\documentclass[12pt,preprint]{aastex}

\slugcomment{ApJ, in press}

\shorttitle{Debris disks in binary systems}
\shortauthors{Trilling et al.}

\begin{document}


\title{Debris disks in main sequence binary systems}


\author{D. E. Trilling\altaffilmark{1}, J. A. Stansberry\altaffilmark{1},
K. R. Stapelfeldt\altaffilmark{2}, G. H. Rieke\altaffilmark{1},
K. Y. L. Su\altaffilmark{1},
R. O. Gray\altaffilmark{3}, C. J. Corbally\altaffilmark{4},
G. Bryden\altaffilmark{2},
C. H. Chen\altaffilmark{5,6},
A. Boden\altaffilmark{2}, C. A. Beichman\altaffilmark{2}}

\altaffiltext{1}{Steward
Observatory, The University of 
Arizona, 933 N. Cherry Avenue, Tucson,
AZ, 85721}
\altaffiltext{2}{Jet Propulsion Laboratory, MS 183-900, California Institute
of Technology, 4800 Oak Grove Drive, Pasadena, CA 91109}
\altaffiltext{3}{Department of Physics
and Astronomy, Appalachian State University,
Boone, NC 28608}
\altaffiltext{4}{Vatican Observatory Research Group,
Steward Observatory, The University of Arizona,
933 N. Cherry Avenue, Tucson, AZ 85721}
\altaffiltext{5}{NOAO, 950 North Cherry Avenue, Tucson, AZ 85719}
\altaffiltext{6}{{\it Spitzer} Fellow}


\begin{abstract}
We observed 69~A3-F8 main sequence
binary star systems using
the Multiband Imaging Photometer for {\it Spitzer}
onboard the {\it Spitzer} Space Telescope.
We find emission significantly in excess of
predicted photospheric flux levels for 9$^{+4}_{-3}$\%
and 40$^{+7}_{-6}$\%
of these systems at 24~and 70~\micron,
respectively. Twenty two systems total have
excess emission, including
four systems that show excess emission
at both wavelengths.
A very large fraction (nearly 60\%) 
of observed binary systems
with small ($<$3~AU) separations have excess
thermal emission.
We interpret the observed infrared excesses as thermal
emission from dust produced
by collisions in planetesimal belts.
The incidence of debris disks around
main sequence A3-F8 binaries is marginally higher than that
for single old AFGK stars.
Whatever combination of nature (birth conditions of 
binary systems) and nurture (interactions between
the two stars) drives the evolution of debris disks in
binary systems, it is clear that 
planetesimal formation is not
inhibited to any great degree.

We model these dust disks through
fitting the spectral energy distributions and
derive
typical dust temperatures in the range 100--200~K
and 
typical fractional luminosities around
$10^{-5}$, with both parameters similar to other
{\it Spitzer}-discovered debris disks.
Our calculated dust temperatures suggest that about half the
excesses we observe are derived from circumbinary
planetesimal belts and around one third
of the excesses clearly suggest circumstellar
material.
Three systems with excesses have dust in 
dynamically unstable regions, and we discuss
possible scenarios for the origin of this
short-lived dust.
\end{abstract}



\keywords{binaries: general ---
planetary systems: formation ---
infrared: stars}

\section{Introduction}

The majority of solar-type and earlier
main sequence stars
in the local galaxy
are in multiple (binary or higher) systems
\citep{dm91,fm92,lada06}.
Planetary system formation is necessarily
more complicated in multiple stellar
systems because of more complex dynamical interactions.
However, protoplanetary disks are known to exist
in pre-main sequence binary systems
both from spectral energy distributions
\citep{ghez93,prato03,monin06}
and from images
\citep{koerner93,stapel98,guilloteau99}.
Some older binary systems also offer evidence
of planetary system formation,
with both planets \citep{patience,egg04,konacki,bakos}
and debris disks \citep{aumann85,patten,koerner2000,prato2001}
known.
Planetary system formation --- broadly
defined --- must be common in a
significant fraction of multiple stellar
systems.


Studying planetary system formation
through direct observation of planets orbiting other
stars is prohibitively challenging at present.
The nearest targets (for which we have the greatest
sensitivity) are generally mature, main sequence 
stars broadly similar to our Sun,
where the signatures of planet formation have long
since been replaced by processes 
endemic to mature planetary systems.
We must therefore study the properties of 
planetary systems indirectly.
It is generally thought that the formation
of planetesimals is a natural byproduct of
(advanced) planetary system formation; our
Solar System's asteroid belt and Kuiper
Belt are remnant small body populations that
reflect the epoch of planet formation.
These small bodies, in our Solar System
and in others, occasionally collide, producing
collisional cascades that ultimately produce
dust. 
Because
dust
is the most easily observable component
of planetary systems due to
its relatively large surface area,
one avenue to understanding planetary system
formation is to study dusty debris disks
around other stars.
However,
no sensitive, systematic examination of the frequency
of debris disks --- signposts of planetary system formation ---
in multiple systems has
been carried out.

Dust heated by stellar radiation
to temperatures of tens to hundreds
of Kelvins
is best observed
at mid- and far-infrared wavelengths, where
the contrast
ratio between the thermal emission of the 
dust and
the radiation of the star
is most favorable.
In many cases the dust temperature and fractional
luminosity can be measured or constrained from
the observations.
Under certain assumptions
of grain properties (size,
albedo,
emissivity, size distribution)
estimates can be made of
the dust mass present
and potentially of the properties
of the planetesimals
that produced the observed dust grains
\citep[e.g.,][]{chas1,vegasu}.

Dust in planetary systems
generally must be ephemeral because the
timescales for dust removal are short
compared to the main sequence ages of the
host stars \citep[e.g.,][]{bp1993}.
The processes of dust production and
removal are more complicated in multiple
systems than around single stars, but
any
dust must nevertheless
be regenerated from
a source population of colliding
bodies.
Dust production can either be through
a continuous collisional cascade, through
stochastic (occasional) collisions,
or derived from individual bodies (e.g., sublimation
from comets).
Ultimately, a relatively substantial population
of larger bodies
(planetesimals: meter-sized up to
planet-sized)
is implied under either
model of dust production, and
argues that
planet formation must have proceeded
to some degree in every system with dust, and therefore
every system with excess thermal emission.

There are extensive programs 
with the {\it Spitzer} Space Telescope
to study
debris disks around single
stars
\citep[e.g.,][]{chas1,chas2,astars,serena,bryden,su},
but binaries --- a majority of solar-type stars ---
have generally been explicitly omitted from these
surveys.
To understand the processes of planetary
system formation and evolution in this
common hierarchical system
we have
carried out a {\it Spitzer} survey for infrared excesses
around 69~binary star systems to look
for thermal emission from dust grains.
Our primary goal is to address 
whether the incidence of debris disks 
in multiple stellar systems is different than
that for single stars.
Here we present our 
24~and 70~\micron\ observations of 
these 69~systems 
and identify excess emission
from a number of them.
We discuss our overall results and individual
systems of note, as well as the dynamical stability
of dust in binary systems.
We conclude with a discussion of 
the implications of our observations for
planet and planetary system formation
in binary systems.

\section{Sample definition}
\label{sample}

We observed 69~binary (in some cases, multiple)
main sequence star systems in order to study the processes
of planetary formation in multiple systems and particularly
to search for effects of binary separation on the presence
of debris disks.
We
chose to observe late A through early F stars for reasons of economy: their
photospheres are bright and we could thus
reach a systematic sensitivity
limit for a significant sample in the shortest observing time.
The primaries in our sample are
18~A~stars (A3 through A9)
and 51~F~stars (F0 through F8).
We have not done an exhaustive study
for higher multiplicity (greater than binarity) for 
the 69~systems in our sample.
Our targets were vetted to eliminate high backgrounds,
and were chosen independent of whether IRAS data implies
any excess for that system.
Our target list also excludes systems with extreme flux
ratios between the two components, and the secondary is generally
G-type or earlier; in practice, this information
is available for only one third of our targets.

Our primary goal is to determine
whether the incidence of debris disks
in binary systems is different than
that for single stars.
Our secondary goal is to determine 
whether there is any effect on debris disk
properties due to binary separation.
Our sample is therefore divided into three subsamples
by binary separation to look for possible
trends in the frequency of infrared excess (that is,
planetary system formation) as a function of binary
separation. (In some cases, these separations are the
projected separations, not the actual orbital
distance.)
21~targets in this program have separations less
than 3~AU;
23~systems have separations
of 3--50~AU; and
19~systems have separations
of 50--500~AU.
Our sample also includes 6~systems
with very large separations ($>$500~AU).
We present results for this last group in this paper,
but do not include them in our
analysis of excess as a function of binary separation.

Typical distances to our targets
are 20--100~pc, though a couple of
systems are as close as 12~pc.
The angular resolution of {\it Spitzer}
24~\micron\ observations is 6\arcsec\
(and 18\arcsec\ at 70~\micron), and almost
all of our systems have angular separations
smaller than this and are therefore unresolved
at both {\it Spitzer} wavelengths.
A handful of systems are resolved (in some
cases barely) at one or both {\it Spitzer}
wavelengths. Our photometric treatment
of both resolved and unresolved systems is
discussed in Sections~\ref{spitzer} and~\ref{modeling}.

The physical properties
of these systems --- including (projected) binary separation and
age, both of which may have an effect on the rate of occurrence of
debris disks --- are reported in
Table~\ref{targetinfo}.
Appendix~\ref{details} gives 
details of our derivations of stellar
properties for this sample.

\section{Observations and data reduction}
\label{observations}

\subsection{{\it Spitzer} observations}
\label{spitzer}

A listing of the 
observations for this program
({\it Spitzer} PID~\#54)
is given in Table~\ref{obsinfo}.
All {\it Spitzer} observations were made
between January, 2004, and March, 2005.
We used the Multiband Imaging Photometer
for {\it Spitzer} \citep[MIPS;][]{mips} to make observations
of each system at 24~\micron\ and, for most
systems, 70~\micron\ (effective wavelengths 23.68~and 71.42~\micron, respectively).
All stars were observed using the MIPS Photometry observing template
in small-field mode. The 24~\micron\ observations were all made
using 3~sec
DCEs
(data collection events)
and a single template cycle. The 70~\micron\ observations typically used
10~sec DCEs and 5 to 10 template cycles.


Data were processed using the MIPS instrument team Data Analysis Tool
\citep{gordon05}. For the 24~\micron\ data basic processing included
slope fitting, flat-fielding, and corrections for droop and readout
offset (jailbar). Additional corrections were made to remove the effects
of scattered light (which can introduce a gradient in the images and
an offset in brightness that depends on scan-mirror position), and
the application of a second order flat, derived from the data itself,
to correct latents that were present in some of the observations. The
70~\micron\ data processing was basically identical to that of the {\it
Spitzer} pipeline (version S13).  Mosaics were constructed using pixels
1\farcs 245 and 4\farcs 925 square at 24~and 70~\micron,
respectively (about 1/2 the
native pixel scale of those arrays).

We used aperture photometry to measure the fluxes from our target systems.      
Aperture corrections were computed using smoothed STinyTim model PSFs           
\citep{krist02} for a 7000~K blackbody source. The model PSFs were smoothed     
until they provided good agreement with observed stellar PSFs, as described     
in \citet{gordon06} and \citet{chad}.

The PSF
full width at half maximum at 24~and 70~\micron\ is 6\farcs 4 and 19\farcs 3,
respectively.
Systems with angular separations less than 6\arcsec\ 
are unresolved at both MIPS wavelengths. For these targets,
fluxes were measured using relatively    
small apertures of 9\farcs 96 and 39\farcs 4 in diameter (at 24~and 70~\micron, respectively)
to improve the signal-to-noise ratio (SNR) of the measurements. 
(In a few cases at 24~\micron\
nearby sources contributed
some flux at the target location, so we used apertures 25\% smaller
than those just described to reduce contamination.)
Systems with angular separations between 6\arcsec\
and 30\arcsec\ are resolved at 24~\micron\ but not
at 70~\micron.
For these cases,
we used                                                           
apertures 35\arcsec\ in radius
to measure the system-integrated flux. Where the components       
were visible and                                                                
clearly separated (at 24~\micron), we compared the photometry from the large aperture with      
the sum of the                                                                  
fluxes from the individual components (measured using the smaller apertures)    
as a cross check.                  
Five systems have large enough angular
separations ($>$30\arcsec) that they are resolved not
only at 24~\micron\ but also at
70~\micron:
HD~142908, HD~61497, HD~77190, HD~196885, and HD~111066.
For these five systems, only photometry for
the primary is measured, modeled, and reported;
we have no measurements for the companions through
either being too faint or out of the field of view.

The photometric
aperture was centered at the center-of-light of each target except
in cases where the 70~\micron\
detection was weak or there was cirrus or background contamination,
where
we forced the aperture to be centered at the target coordinates. 
The fluxes we report are based on conversion factors of 
1.048~$\mu$Jy/arcsec$^2$/(DN/s) and 16.5~mJy/arcsec$^2$/U70 at 24~and
70~\micron, equal to the calibration in the {\it Spitzer} Science Center
pipeline version S13
(further details on calibration can be found in \citet{riekecalib}, \citet{gordon06}, and
\citet{chad}).

The 24~and 70~\micron\
{\it Spitzer} photometry for all sources
observed in this program is
reported in Table~\ref{photom}, together with
the system-integrated V~and K~band magnitudes used in photospheric
model fitting (Section~\ref{modeling}).
All targets were strongly detected at 24~\micron,
with intrinsic S/N in the hundreds to thousands.
The 70~\micron\ observations were planned such
that the predicted combined photospheric flux
from the system could be detected with S/N of at least~3
in 1000~second;
the 16~systems
that did not meet this criterion
were not observed at 70~\micron.
We also discard from our statistical sample
the three sources that were observed at 70~\micron\
but not detected,
leaving 50~good observations at 70~\micron.

All measurements are subject to both photometric (measurement)
error and a uniform calibration
uncertainty of 4\% at 24~\micron\ and 8\% at 70~\micron\
\citep{gordon06,chad}.
These
two sources of error are
RSS-combined to calculate the total errors presented in Table~\ref{photom}.

\subsection{Submillimeter observations}

We observed 13 of our systems at 870~\micron\ with the Heinrich Hertz Submillimeter
Telescope on Mt.\ Graham, Arizona.
The data were reduced using the NIC package, which produces mosaicked
images from the 19-channels of the detector, subtracts the ``off'' images
from the ``on,'' and accounts for atmospheric opacity (which we measured
regularly using sky-dips). Flux calibrations were derived from observations
of the planets (primarily Neptune and Mars).
The typical 3-sigma sensitivity
achieved in those observations was $\sim$30~mJy.
None of the 13~systems were detected
above the 3-sigma level, and upper limits for each
system are given in Table~\ref{photom}.
Our submillimeter observing program was 
cut short due to the failure of the 
facility bolometer array,
and the remaining systems
have not been observed by us in the submillimeter.

Assuming an excess temperature of
50~K, the ``minimum temperature fit'' that
we employ below and which gives the maximum
submillimeter flux,
the ratio of 70~\micron\ flux to
870~\micron\ flux is~$\sim$12, so the 
70~\micron\ flux ideally would have to be
greater than $\sim$350~mJy for us to have made
a significant detection in the submillimeter.
HD~13161 is the only target in our sample
with a 70~\micron\ flux greater than 150~mJy.
Since this target unfortunately was not
observed
before the demise of the bolometer array,
it is not surprising that all of our 870~\micron\
observations are upper limits.

For all 13~sources observed 
at 870~\micron\ the 
upper limits do not significantly constrain
the debris disk models
that we present in this paper.

\section{Results}
\label{results}

\subsection{Modeling photospheric fluxes}
\label{modeling}

From published visible and near-infrared data,
we determine the best-fit Kurucz model spectrum;
details of this process are described in
Appendix~\ref{details}.
Many of our systems are resolved in
visible and near-infrared data, but
almost all are unresolved at one or 
both {\it Spitzer} wavelengths (Section~\ref{spitzer}). Our
approach is therefore to combine fluxes
at any wavelength where the components 
are resolved into a single system-integrated
flux measurement (with the five exceptions
listed in Section~\ref{spitzer} and Table~\ref{photom}).


We model the combined flux from each binary system
as a single stellar source. This approach is satisfactory
regardless of the (dis)similarity between the two
spectral types:
for every primary star presented here,
no secondary spectral type changes the
the slope of the Rayleigh-Jeans part of the
spectral energy distribution (SED) by
more than 1\%
from the trivial case of primary and secondary
stars having identical spectral types.
The errors in our predictions
are therefore always small
compared to other sources of error.

Using the best-fit Kurucz
model, 
we predict the fluxes for the {\it Spitzer}
observations at the 
24~and 70~\micron\ effective wavelengths.
These predicted photospheric fluxes are listed
in Table~\ref{photom}.

%
%

In Figure~\ref{sedfig} we show SEDs
for two binary
systems with no excess emission in our MIPS
observations.
These systems 
are representative of our method of
photosphere modeling
and predicting 24~and 70~\micron\ photometry.
The
measured {\it Spitzer} photometry
falls quite close to the predicted fluxes
in all cases.
It is clear that our 
technique of fitting a single temperature
model works quite satisfactorily both in
the visible/near-infrared and also at {\it Spitzer}
wavelengths.

\subsection{Determination of excesses}


We use the ratio (R) of observed flux (F) to predicted flux (P)
to determine the excess threshold and to identify
excess emission.
In Figure~\ref{fp} we show histograms of R24 and R70
for all observed systems.
The R24 distribution
is
well fit by a gaussian centered at R24~=~0.99 
and $\sigma=0.05$.
We 
take a conservative approach,
adopting an excess threshold ratio of~1.15
(Figure~\ref{fp}, top), which
formally is slightly more than 3-sigma.
The R24 dispersion less than unity, which should represent
excursions $\lesssim$3$\sigma$, extends
smoothly down to~0.85, confirming $\sigma=0.05$.

It is more difficult to produce a well-fit gaussian
to the 70~\micron\ data because there are only 50~measurements
(fewer than
at 24~\micron),
of which
more than one third
likely have excesses (Figure~\ref{fp}, middle).
The scatter in the R70 distribution
implies $\sigma=0.10$, centered near unity,
suggesting that we adopt a
3-sigma error threshold of~1.30.
We note that both the R24 and R70 excess thresholds
are consistent with, though perhaps somewhat larger
than, the errors due to 
systematic calibration uncertainties,
giving us confidence that our thresholds
are accurate but also conservative.


In most cases,
our observed 24~and 70~\micron\ fluxes are within 1~sigma
(5\% and 10\%, respectively)
of the predictions (Table~\ref{photom}),
confirming
that our photospheric predictions are good.
Occasionally the 
measured fluxes are less than the predictions by 2--3~sigma,
and a number of cases have observed fluxes
that are greater --- in some cases, substantially
so --- than the predictions.
Some of these individual cases with significant excesses
are discussed in Section~\ref{interest}.
We note, however, that one 24~\micron\ and three 70~\micron\
measurements have R values that deviate from unity by
more than 3$\sigma$
(Table~\ref{photom}).
The existence of these low R24 and R70 values may indicate
that we have underestimated the scatter
in the data, as we would expect no values 
more than 3$\sigma$ below unity for a sample 
of this size. This may in turn
imply that a few systems
that we identify as excesses based on their
R values
may
be spurious (noise
rather than true excesses). For this reason, we introduce
the additional requirement of having significant excess emission,
as follows.


%

%

We calculate the significance ($\chi$) of a
detected excess as 

\[
\chi = \frac{F - P}{\sigma}
\]

\noindent where F and P are as defined above
and 
$\sigma$ is the total error (photometric error [noise]
and calibration error, added in quadrature)
of the measurement.
This figure of merit $\chi$ is calculated for
each measurement at each wavelength (Table~\ref{photom}).
The significance of a measurement that
exactly matches the prediction is zero.

Formally, 
to identify excess emission from a system, we 
require that R
be greater than the thresholds derived above
{\em and} that the significance be 2.0~or greater. 
Thus, systems like HD~8556
are, sensibly, excluded from being valid
excess detections (with R70~=~1.35 and
$\chi_{70}=1.01$, the ``excess'' $F-P$ here 
is comparable to the total error $\sigma$).
We list all valid excess systems in Table~\ref{excesstable}.

\subsection{Overall results}


Using the criteria explained above,
we find that
9$^{+4}_{-3}$\% of the systems
have excess emission at 24~\micron\ (6/69)
and
40$^{+7}_{-6}$\%~show
excess at 70~\micron\ (20/50),
using binomial errors that include 68\% of
the probability (equivalent
to the 1$\sigma$ range for gaussian errors),
as defined in
\citet{burg03}.
Four systems have excesses at both wavelengths.

Individual 24~\micron\ excesses range from 16\% to
47\% above the predicted combined photospheric flux (Figure~\ref{fp},
Table~\ref{excesstable}).
R70 ranges from~1.3
to more than~25 (Figure~\ref{fp},
Table~\ref{excesstable}).
%
%
%
A 100\% excess above
the predicted combined photosphere (that is, R~of~2)
means
that the thermally emitting dust in the system
is as bright as the total flux from the star at the specified
wavelength. Eleven systems
have R70$\geq$2.0.
These very large excesses indicate
relatively high fractional luminosities,
which in turn imply large amounts of dust
in these systems.

\subsection{Identification of false IRAS excess}

Seven sources have IRAS 25~\micron\ fluxes \citep{fsc} more
than 30\% above the predicted photosphere:
HD~13161,
HD~13594, 
HD~16920,
HD~20320,
HD~80671,
HD~83808, and
HD~118216 (considering only quality flag~3 data).
One source has a measured IRAS 60~\micron\ flux
more than 50\% above its predicted photosphere:
HD~13161 (again, considering only quality flag~3 data).
Except for HD~13594, 
the IRAS measurements, after color- and 
wavelength-corrections, are all quite
consistent with our {\it Spitzer} measurements
(Table~\ref{photom}, Table~\ref{excesstable})
and we confirm the IRAS-detected excesses for
these six systems 
(indeed, as indicated
in Section~\ref{interest} and Appendix~\ref{notes},
several systems were identified as excess
systems previously based on the IRAS data).
In contrast,
the color-corrected IRAS 25~\micron\ measurement for
HD~13594 is 50\% higher than 
our 24~\micron\ measurement, and two sigma
above the predicted photospheric flux at
25~\micron\ (using the IRAS reported error).
We see no additional sources in our 24~\micron\
image of this target; 
contamination in the large IRAS beam is probably
not the explanation for the high 25~\micron\
flux. 
We
speculate that the
IRAS measurement is simply anomalously high.
Since debris
disk searches are often still based on catalogs
of IRAS-selected excesses, 
we identify here HD~13594 as a false excess
so that future disk searches need not 
spend time observing this source.

\section{Analysis of observational results}
\label{analysis}

\subsection{Excess as a function of binary separation}

R24 and R70
as a function of separation
are shown for the individual measurements in
our sample
in Figure~\ref{separations}.
The systems with excess are shown in Figure~\ref{hist}.
The excess rates for small and large separation
systems are around 50\%,
and 
there are fewer medium separation systems
with excesses than either small or large separation
systems.
The relative lack of excesses in systems with
medium separations
confirms our theoretical expectations
(see Section~\ref{bias}).
A smaller excess rate for medium
separation systems is also in agreement
with observations of pre-main-sequence
binaries that suggest that
systems with separations
1--50~AU (approximately equal to our 
medium separation bin)
have significantly fainter disks
than systems with large separations
\citep{jensen94}.
The significances of high excess rates for
small and wide separation binaries, and a
low excess rate for medium separation binaries,
are discussed in Section~\ref{context}.


\subsection{Properties of the dust disks}

\subsubsection{Dust temperatures}

In previous sections we have discussed excess emission
detected
at 24~and 70~\micron.
We now move to the astrophysical interpretation of this
excess emission as thermal radiation from dust grains
heated by the radiation fields of the star(s).
We assume these grains are large and model them as blackbodies.
(We briefly explore the implications of non-black-body grains
in Section~\ref{nonbb}.)
We derive best-fit temperatures
for these dust grains, assuming a single temperature for the
ensemble population, based on a uniform distance
from a single radiation source.
While the radiation and temperature
fields in binary systems are certainly more complicated
than these simple assumptions,
in general the results of these
approximations will be adequate to help us
understand the properties
of the systems we observe and allow comparisons among
the systems presented here, and to results presented elsewhere.

To calculate dust temperatures for systems with
excesses, we used the following techniques.
For systems with both 24~and 70~\micron\ excesses
(four systems), we fit a blackbody
to the
excess emission in both bands. For systems with only a
70~\micron\ excess, we fit
the 70~\micron\
excess emission and the 3-sigma upper limit on the 24~\micron\ emission
(that is, the predicted flux plus three times the
1-sigma error bar). This
approach
produces an upper limit to the excess temperature (and dust luminosity)
consistent
with our data. For the three systems with excess emission detected only at
24~\micron, we took the approaches
described in Section~\ref{interest}.

Our calculated temperatures for excesses are listed in
Table~\ref{excesstable}.
We again emphasize that these temperatures
are the {\em maximum} temperatures that can be
fit to the SEDs.
In Section~\ref{mindust} we explore
``minimum'' temperature solutions
for the excess systems, and the implications thereof.



\subsubsection{Dust distances}
\label{dustdist}

After solving for the dust temperature ($T_g$), we can
calculate the orbital distance $r$ (in AU) of the dust
through equation~3 from \citet{bp1993}:

\[
r = \left(\frac{278}{T_g}\right)^2 \left(\frac{L_\star}{L_\odot}\right)^{0.5}
\]

\noindent where $L_\star$ is the (combined) stellar
luminosity.
We calculate the combined stellar luminosity of the 
host star(s) simply through

\[
L_\star = 4\pi R_\star^2 \int{{\rm Kurucz~model}} = 4\pi R_\star^2 \sigma T_{eff}^4
\]

\noindent where $R_\star$ is 
the stellar radius \citep[from][]{drilling}
and
the stellar effective temperature
$T_{eff}$ is given in Table~\ref{targetinfo}.

Because we know the luminosity of the (combined)
host stars from our photospheric fitting, for each system
we can
also calculate the (single) radial distance of the
source of the excess;
these distances are reported in Table~\ref{excesstable}.
Because the temperatures we use are the maximum temperatures,
the distances derived in this way are {\em minimum} distances.
This logic of assigning a single distance to the dust,
based on the combined stellar luminosity, is obviously
the simplest possible model.
Many more complex geometries and solutions
are possible; we discuss some of these in
Section~\ref{unc_dust}.

\subsubsection{Fractional luminosities}

The fractional luminosity of a dusty debris
disk is the ratio of the integrated luminosity of the
emission by the dust to the integrated luminosity
of the host star(s). The former
is the blackbody fit to the excess(es),
as described above, and the latter is
the best-fit Kurucz model described
in Section~\ref{modeling}.
The fractional luminosity can be understood
visually from the SEDs shown in Figure~\ref{sedfig}.

We derive fractional luminosities for each
system; 
because we fit maximum temperatures to the infrared
excesses, the derived fractional luminosities
are also maximum values.
We explore in Section~\ref{mindust} the impact of the ``minimum'' temperature
alternate assumption on fractional luminosity.
We list the derived fractional luminosities for
all 22~systems with formal excesses
in Table~\ref{excesstable}; these fractional
luminosities are plotted in Figure~\ref{acrit}.
Systems with excesses at 24~\micron\ generally have
large fractional luminosities because the
dust temperatures are warm.
This can be seen in Table~\ref{excesstable}, where
the largest fractional luminosity is for
HD~83808,
the system with the highest excess temperature.
We discuss the details of that system in the following section.

Our fractional luminosities are mostly in the
range $10^{-5}$ to $10^{-4}$ (Table~\ref{excesstable} and Figure~\ref{acrit}).
This range is consistent with values found
by other surveys of debris disks around ``old'' stars
(as described in Section~\ref{comparison}).
We show in Figure~\ref{acrit} fractional
luminosity as a function of dust distance
in units of binary separation; there is no obvious
trend.
There is an apparent limit near $10^{-5}$, which
is approximately the MIPS detection threshold
\citep[see also][]{bryden}.

%

Our observations at 24~and 70~\micron\ 
place no constraints on colder ($\sim$30~K) 
disk components. Our 13~submillimeter upper
limits preclude the existence of very massive
cold disks but place no useful constraints on 
modest (fractional luminosity $\sim$10$^{-5}$) cold disks.
Although we 
refer to the fractional luminosity values we 
derive as maximum values, it is 
worth noting that a cold,
massive disk component could exist for almost all
systems in our sample. This putative cold disk
could imply a larger fractional luminosity than
the ``maximum'' values that we report here.

\subsection{Systems of interest}
\label{interest}

\begin{figure}
\vspace*{-32mm}
\epsscale{.7}
\plotone{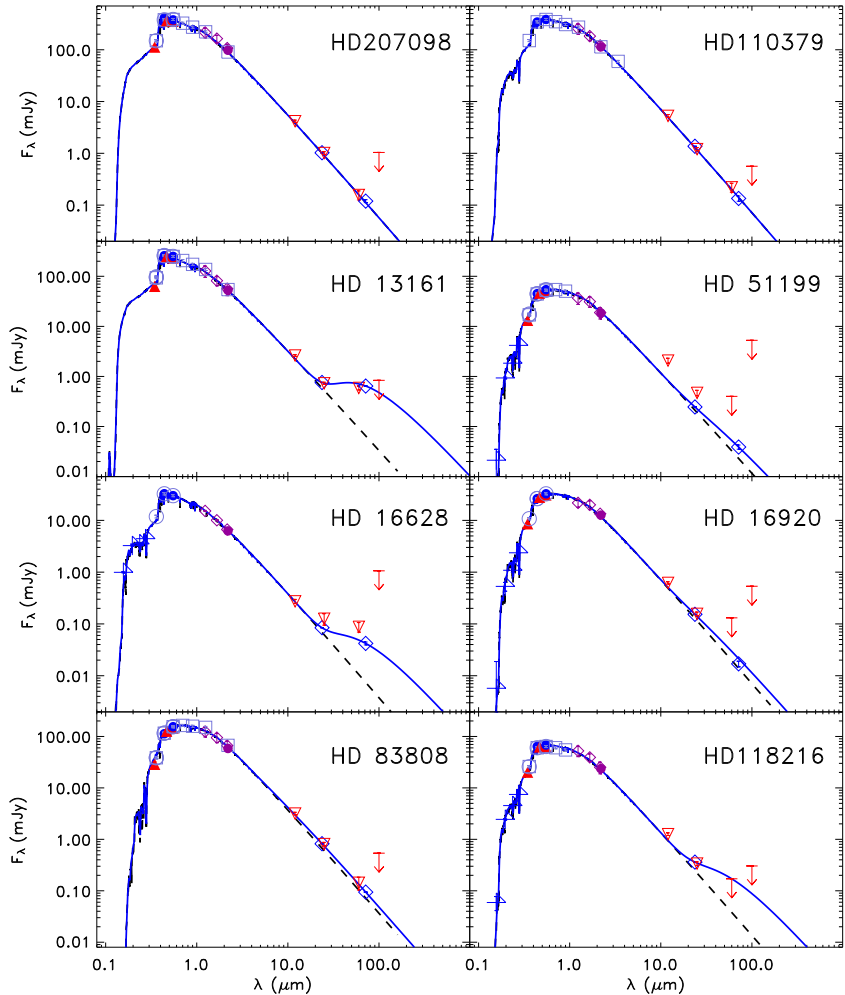}
\epsscale{1}
\vspace{-3ex}
\caption[]{{Spectral energy distributions (SEDs) for eight binary systems.
The top two panels show systems that have no
excess emission at either {\it Spitzer} wavelength,
whereas the remaining six panels show SEDs for
systems with various kinds of excesses, as
described in Section~\ref{interest}.
In all cases, 
the blue solid line shows our
best-fit SED, and the dashed black
line is the best-fit
scaled Kurucz
model spectrum that is our modeled photospheric
flux.
The Kurucz model is fit
to the combined flux of the two stars,
using
optical
and near infrared data:
2MASS data 
is shown as purple diamonds, with ``super-K''
(see Table~\ref{photom}) as a filled purple
circle, and various other ground-based visible
datasets (see Appendix~A) are shown.
Our {\it Spitzer}/MIPS measurements are shown
as blue diamonds, with
IRAS data points are shown as red downward pointing
triangles (or upper limits). 
In some cases,
IRAS fluxes are significantly above our 
best-fit SED due to flux from
other (non-targeted) sources in the IRAS beam
that are excluded by our aperture photometry.
Since IRAS data is not used in determining
the best Kurucz model or the best-fit SED,
these discrepancies do not affect our analysis
or interpretation.
The top two panels demonstrate that
our 24~and 70~\micron\
photospheric predictions
generally correspond quite well to predicted fluxes
for systems with no excesses.
Details of the best-fit
disk models are presented in the text and in
Table~\ref{excesstable}.
Fractional luminosity is the ratio
of the total emission from the disk to the
total emission from the star(s).
For HD~13161, HD~51199, HD~16628, and HD~83808,
the SEDs shown here correspond to the debris
disk solutions given in Table~\ref{excesstable}.
For HD~118216, we show here an SED that corresponds
to a dust temperature of 134~K (see Section~\ref{interest}).
For HD~16920, we show here an SED that corresponds
to a dust temperature of 260~K (Section~\ref{interest}).
}}
\label{sedfig}
\end{figure}

\begin{figure}
\includegraphics[angle=270,scale=0.60]{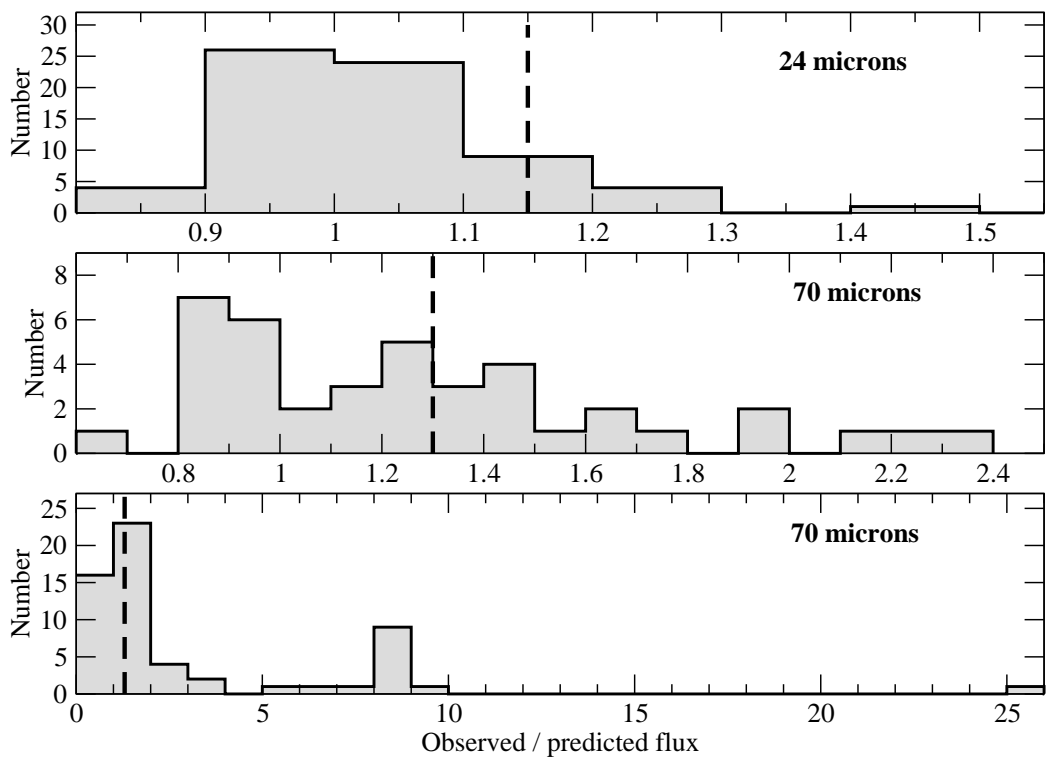}
\caption[]{Histogram showing
R24~(upper) and R70~(middle and lower),
where R is the ratio of observed flux
to predicted flux.
The middle and lower panels show the same
data, but with different bin sizes and horizontal scales
in order to show both the inner core (R70 near unity;
middle panel) and the total range of R70 (lower
panel). (Some R70 values low and high are beyond 
the compressed range of the middle panel but
appear in the bottom panel.)
There are 69~systems with 24~\micron\ measurements
and 50~systems with 70~\micron\ measurements.
The excess threshold values of 
1.15 (24~\micron) and 1.30 (70~\micron)
are indicated with dashed lines.
}
\label{fp}
\end{figure}

\begin{figure}
\includegraphics[angle=270,scale=0.60]{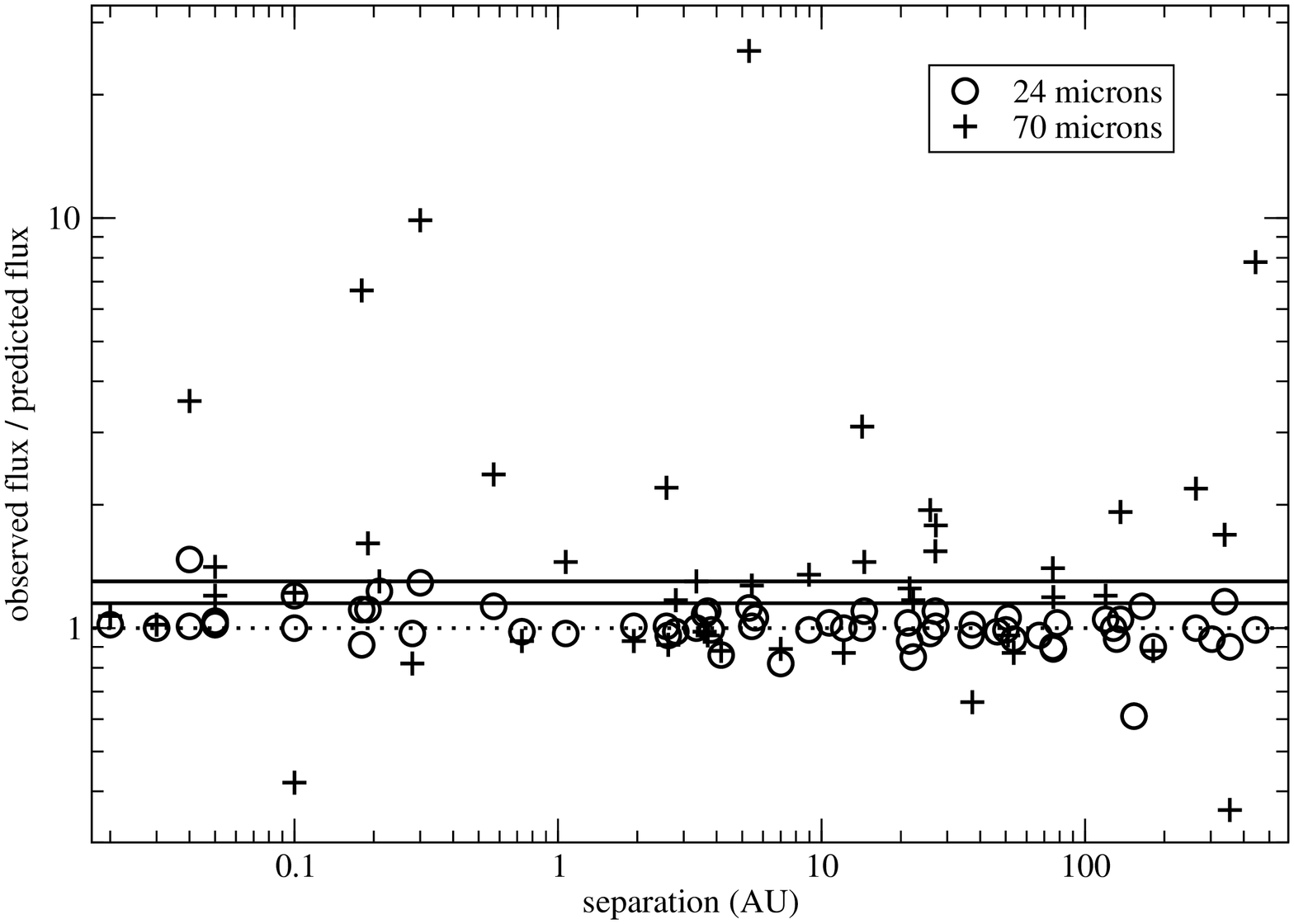}
\caption[]{Ratio of observed to predicted
fluxes for all systems as a function
of binary separation.
Circles show
data for 24~\micron\ observations and
crosses show results for 70~\micron\ observations.
In all cases the errors in R24 and R70
are smaller than the symbols.
The solid horizontal
lines show our criteria for identification
of excess at
R24~=~1.15 (lower line) and
R70~=~1.30 (upper line), and the
dotted line shows R~=~1.00, for
guidance.
There is no significant trend with separation,
although systems with separations 1--10~AU may
have fewer, or smaller, excesses.}
\label{separations}
\end{figure}

\begin{figure}
\includegraphics[angle=270,scale=0.60]{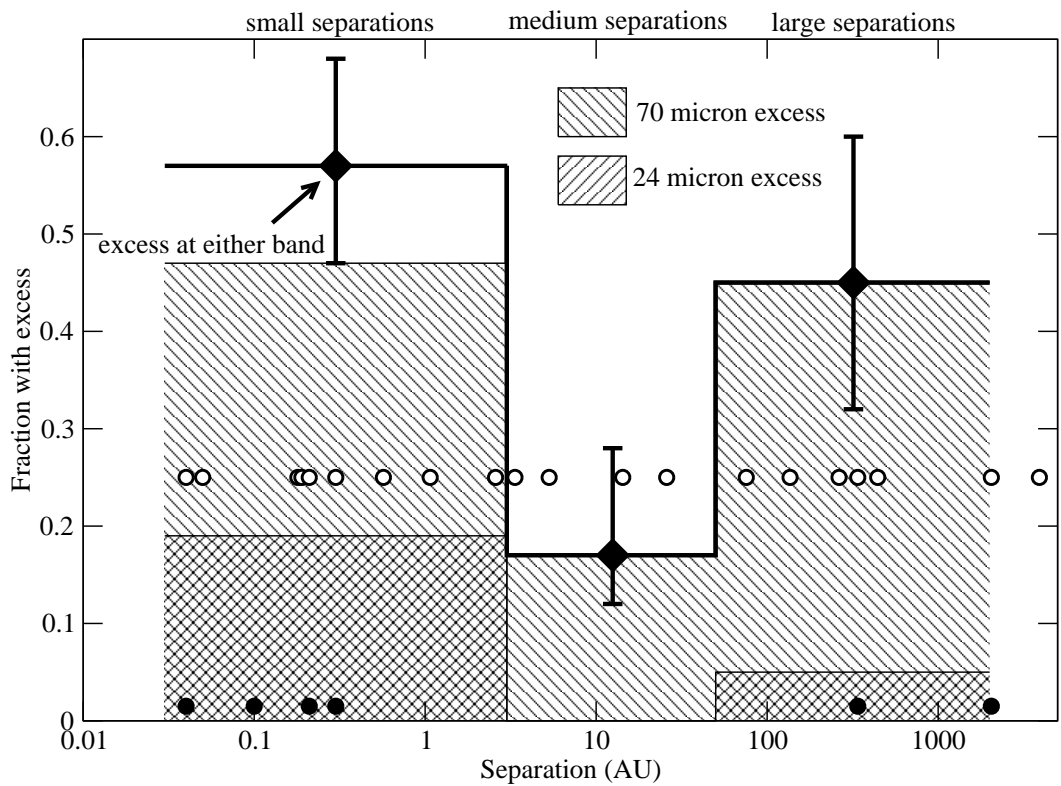}
\caption[]{Fraction
of binary systems in
each of three logarithmic bins
(0--3~AU; 3--50~AU; 50--200~AU) that
have 24~\micron\ (diagonal pattern
lower left to upper right), 70~\micron\ (diagonal
pattern lower right to upper left),
or 24-or-70~\micron\ excesses (clear).
Binomial error bars are shown for the
24-or-70~\micron\ excesses category.
Each category reads from the bottom
of the plot (that is, the fraction of close
binaries with 70~\micron\ excesses is 47\%).
Some systems have excesses at both wavelengths,
and the number of observed systems is not the 
same at 24~and 70~\micron,
so the combined fractions do not simply
equal the sum of the two subcategories.
The separations of the individual systems
with excesses
contained within each bin are indicated by
the filled (24~\micron) and open (70~\micron) circles
(with arbitrary y-axis values).
Intermediate separation systems
have fewer excesses than small or large separation
systems, as expected (Section~\ref{bias}).
}
\label{hist}
\end{figure}

We show six systems of particular interest
in Figure~\ref{sedfig},
and discuss them here.
These systems present the most interesting
and illustrative cases for SED fitting.
An additional 13~systems of note are discussed
in Appendix~\ref{notes}.

\vspace{1ex}

\noindent {\bf HD~13161, HD~51199, and HD~16628.}
These three systems all have formal excesses at
both 24~and 70~\micron\ (Figure~\ref{sedfig}).
We therefore have a very good measurement of the color
temperature of the excess. HD~13161
was identified as Vega-like --- meaning
likely possessing a debris disk ---
by \citet{sadakane} as well as a number
of later workers, based on IRAS fluxes.
HD~51199 has a relatively strong excess at
24~\micron\ and a relatively weak excess at
70~\micron, implying a somewhat high dust temperature
of 188~K.
The 70~\micron\ flux for HD~16628
is more than five times brighter than the expected
photospheric flux. 

\vspace{1ex}

\noindent {\bf HD~83808.}
HD~83808 also has formal excesses at both
bands (Figure~\ref{sedfig}).
This two-band fit gives an excess temperature of
815~K;
comparable excesses at 24~and 70~\micron\ (23\% and 30\%,
respectively) indicate
that the excess
color is only slightly redder than the star(s), implying
a relatively high excess temperature.

Because this temperature is quite high compared to a
typical dust disk result in this program, we use IRAS
measurements for a consistency check.
The color-corrected IRAS fluxes are
3500, 800, and 130~mJy at 12, 25, and 60~\micron,
respectively; this implies 24~and 70~\micron\ fluxes
of 870~and 96~mJy, respectively
(after scaling by
$\lambda^2$). (We note that the IRAS 60~\micron\
measurement formally has quality flag 1, meaning
an upper limit, but the measurement
is consistent with our higher S/N observation.)
Our 24~and 70~\micron\ measurements
of 822~and 94~mJy, respectively, match the IRAS
data quite well. We can extrapolate our 24~\micron\
measurement to 12~\micron\ (again scaling
by $\lambda^2$), and get 3500~mJy, again matching
the IRAS measurement.
We conclude that the IRAS data are consistent with
our measurements.

Now we look again at the IRAS data for confirmation
of our excess temperature.
The predicted photospheric fluxes (combining
IRAS and MIPS data) at $\lambda$ = [12,24,25,60,70]~\micron\
are [2600, 670, 600, 103, 72]~mJy (after color correction).
The observed fluxes are [3500, 822, 800, $<$130, 94]~mJy
(after color correction). This implies a 12~\micron\ excess
of 900~mJy. An 815~K dust population would imply
an excess at 12~\micron\ of around 400~mJy, for a total
(color-corrected) 12~\micron\ flux of $\sim$3000~mJy.
Since the
error on the IRAS 12~\micron\ data is 6\%, the measured
IRAS 12~\micron\ flux of 3500~mJy is consistent at
a two sigma level with 
emission from an 815~K disk, and may even 
imply a
hotter temperature for the excess.
We therefore feel confident in the
determination of an 815~K excess for this system,
based on detections at both MIPS bands
and an
IRAS 12~\micron\ excess.
This temperature is one of the hottest
debris disk temperatures known.

We note that the primary of HD~83808 has moved 
off the main sequence (see Section~\ref{youth} and \citet{hummel}).
This should not have any significant effect on our
SED fitting or analysis
of excess emission from this system, but adds to the complexity of this system.
For example, it is possible that this giant star
could be ejecting dust, and that the hot excess we
observe may represent a dust shell rather than
a debris disk.
We also note (from Simbad) the presence of a radio source and
an X-ray source within about 15\arcsec\ of HD~83808;
if these sources are related to the HD~83808 system
then further complexities may be implied.

\vspace{1ex}

\noindent {\bf HD~118216.}
This system shows a 47\% excess at 24~\micron\ (Figure~\ref{sedfig}).
We did not observe this system at 70~\micron\ because
its predicted 70~\micron\ flux suggested that, in the
absence of any excess emission, we would not have detected
the photosphere at greater than 3$\sigma$ precision.
The single bandpass (24~\micron) detection unfortunately
does not allow
us to constrain the system's SED,
shown
in Figure~\ref{sedfig}, or the
color temperature of the excess, and we must
turn elsewhere to improve our understanding of
this excess emission.

The IRAS 12~\micron\ flux measurement, after
color correction, is greater than the predicted
12~\micron\ photospheric flux by about
1.7~sigma (measurement error, not including
calibration error). One approach would be to 
assume that this 12~\micron\ measurement is indicative
of excess flux; that logical path implies a maximum
excess temperature around 850~K. However, we are
reluctant to place too much emphasis on a 
1.7~sigma ``excess'' and look elsewhere for
additional constraints.

In Section~\ref{mindust} we argue that the minimum
reasonable dust temperature is 50~K for all systems.
Hence, we calculate the fractional luminosity 
for 50~K dust in this system by forcing
the emission from the dust to pass
through the measured 24~\micron\ flux value.
In doing so, we find
that the hot (here, 850~K) solution has a lower fractional
luminosity than the cold (here, 50~K) solution,
in contrast to the pattern typically seen for debris
disks detected in this program (Figure~\ref{acrit}).
Because
of our lack of good constraints for this system,
we present this minimum (and hence conservative)
solution in Table~\ref{excesstable} and Figure~\ref{acrit}
as our ``best-fit'' solution.
As usual, Figure~\ref{acrit} shows the best-fit solution
(here, 50~K)
as a symbol, with a tail extending to the extreme
other solution (here, 850~K, from the above analysis).
A dashed line is used for
this tail to indicate that more assumptions than
usual were made for this system.
Note that, in comparison for most other systems with
excesses in our sample, for HD~118216 we derive
a {\em lower limit} on the temperature.



As a consistency check, we use the 
IRAS 60~\micron\ upper limit of 171~mJy \citep{fsc} to derive a temperature
for the excess. We subtract the predicted photospheric
flux at 60~\micron\ and ascribe the difference from the
upper limit (133~mJy) as all due to potential excess.
Using this 60~\micron\ ``excess'' and our 
24~\micron\ (MIPS) measurement we find a color
temperature of 134~K.
To pick a representative temperature, we show
this 134~K fit in Figure~\ref{sedfig} (but use
the bounds 50--850~K
elsewhere in this paper).
This temperature falls within the 
bounds presented above of 50--850~K, and so
is consistent with our bounds given above,
but it is clear that
further data on this target would help eliminate
the need for
some of the above assumptions and better constrain
the dust temperature.

\vspace{1ex}

\noindent {\bf HD~16920.}
For this system we formally detect an excess
at 24~\micron\ and formally do not detect
an excess at 70~\micron, where the excess
ratio of 1.22 does not meet our excess criterion and where
$\chi_{70}$ of~1.22 may indicate that
there is no significant excess for this system at
70~\micron\ (Figure~\ref{sedfig}).
The IRAS data are consistent with our observations.

Because we have data at both MIPS wavelengths,
here we can follow the process
described above for the case of a clear excess
at one wavelength and no excess at the other wavelength.
We fit a blackbody to the measured
24~\micron\ data (where the excess is found) and the predicted
70~\micron\
photospheric flux plus three times the error at
70~\micron\ (see Table~\ref{photom}).
For an excess at 24~\micron\ and no excess at
70~\micron, this technique gives a {\em minimum}
temperature for the excess (through an upper limit
on the 70~\micron\ flux).
This minimum excess temperature is 260~K.
With no evidence of excess in the IRAS
12~\micron\ data, we must additionally
make a ``maximum temperature'' assumption.
We choose
500~K, which is similar to the temperature derived
for the dust in HD~69830 \citep{chas1} and is warmer
than almost all derived debris disk temperatures
from this and other programs.
We solve for the fractional luminosities for these
bracketing temperatures, again forcing the derived
excess emission profile to pass through the
observed 24~\micron\ point.
We use the 260~K solution as our best fit,
shown in Table~\ref{excesstable}, Figure~\ref{sedfig},
and Figure~\ref{acrit}.

It is unusual in our MIPS
surveys for
debris disks to show formal excess at 24~\micron\ and not at 70~\micron\ (for
systems strongly detected at both wavelengths).
The best example of a system with that unusual
excess pattern is HD~69830, which \citet{chas1}
interpret as hot dust created in a very recent
asteroidal or cometary collision. We argue below that at
least two binary systems with excesses also show
evidence of recent dust-producing collisions.
Note that HD~69830 was recently found to have 
three planets orbiting that star \citep{lovis}, perhaps further
linking 24~\micron\ only excesses with planetary
system formation.


\section{Dynamics: Where is the dust?}
\label{dynamics}

\subsection{Stability in binary systems}
\label{unstable}

%
%
%
\citet{hw99} carried out a study
of the stability zones in binary systems by placing
test particles in binary systems with a range of
mass ratios and eccentricities;
in all cases the test particles have
zero eccentricity.
(A similar stability study was carried out by \citet{verrier}
for the planet-bearing binary system $\gamma$~Cephei.)
Holman \& Wiegert derive a ``critical semimajor axis,'' $a_c$,
which is the maximum (for circumstellar
material) or minimum (for circumbinary
material) semi-major axis 
where the test particle is stable 
over $10^4$~binary periods, though
resonances
may reduce stability even in regions safely
beyond $a_c$.
For circumstellar material,
the critical radius
within which material is stable
is typically
0.1--0.2~in
units of the binary separation;
for circumbinary material,
the critical radius outside of which
material is stable
is typically
3--4~in
the same units (Figure~\ref{schematic}).
The locations of these stability boundaries
change for various
binary eccentricities and mass ratios --- information
that is 
known for some of the binary systems
with excesses (see Table~\ref{excesstable}) ---
but these changes are relatively small
for most of the binary systems we consider.
\citet{whitmire}, in
a similar study to \citet{hw99},
found an unstable zone that is somewhat
broader (that is, a greater range of orbital
distances that are unstable), as did
\citet{moriwaki} and \citet{fatuzzo}.
To be conservative in our identifying
dust in unstable disks (that is, to identify
a lower limit on the number of unstable disks),
we adopt the \citet{hw99}
criteria, and note that 
using the broader instability zones would 
increase the number of systems with 
dust in unstable locations.
As a caveat, we note that the Holman \& Wiegert stability
criteria only apply to test particles that experience
no non-gravitational forces, and therefore do not explicitly
apply to dust. However, their stability arguments
do apply to the asteroidal bodies that collide
to produce dust (Section~\ref{belts}).

\subsection{Debris disk geometries in our sample}

\begin{figure}
\vspace*{-20mm}
\includegraphics[angle=270,scale=0.60]{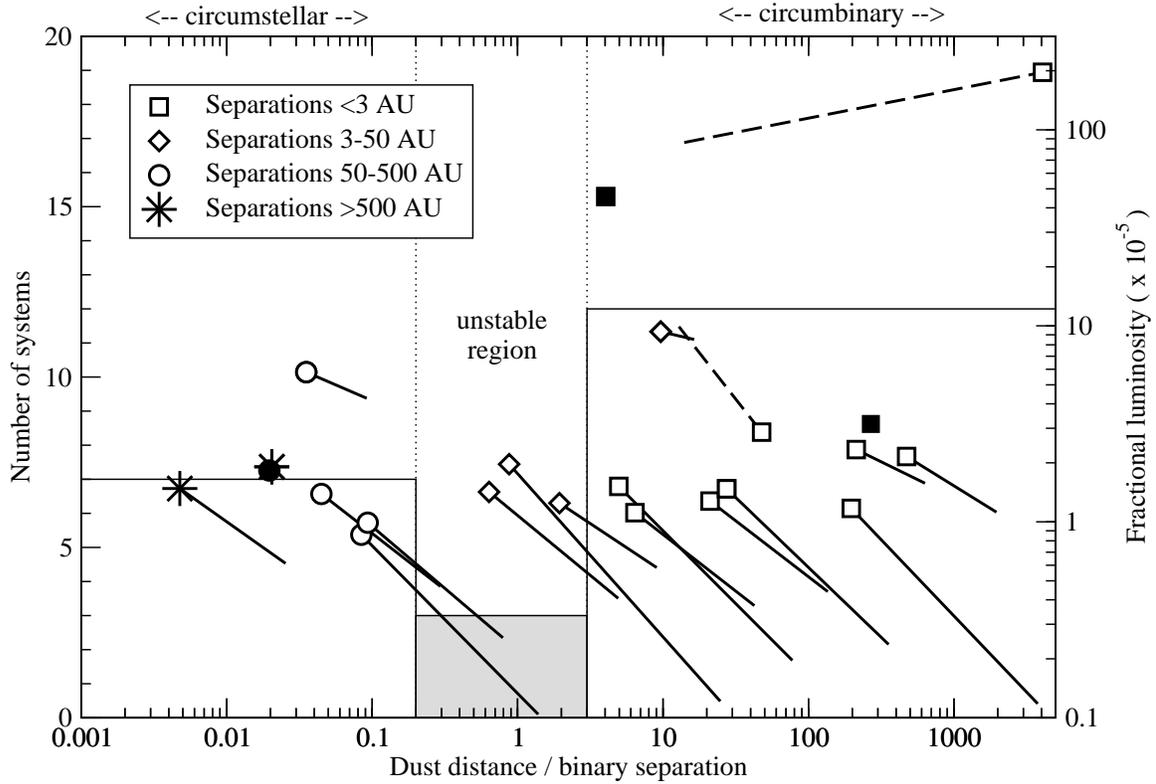}
\caption[]{Histogram of dust distance
in units of binary separation (left axis)
and fractional luminosity as function of
dust distance in units of binary separation (right axis).
{\it Left axis:} The dashed vertical lines show the
approximate boundaries of the unstable
zone (histogram bar shaded grey).
Dust in two systems is
found to reside within this dynamically
unstable region (Table~\ref{excesstable}).
{\it Right axis:} There
is no strong trend
between fractional luminosity and dust location.
Binary systems with small, medium, large, and very large physical
separations are indicated.
Not surprisingly, circumbinary disks are generally found
in small separation systems and circumstellar disks are
found in large separation systems. Dust
in unstable regions is found only in
separation systems, as expected
(Section~\ref{bias}).
Fractional luminosities for the maximum temperature 
cases
are indicated by the symbols.
``Tails'' on the symbols indicate the locus
of solutions, from maximum temperature solutions (symbols) to
50~K (minimum reasonable) solutions at the other ends of
the tails. (There are two exceptions, where
we instead use the cool solutions as our best solutions.
The range of solutions for these two systems
extends to the left in this plot, as described
in Section~\ref{interest}.
We show those ranges as dashed lines
because we have no good upper bounds for these
systems.) 
The range of solutions generally is not large
enough to change our dynamical classifications substantially.
The four symbols without tails 
(two filled squares, one filled circle, and
one star)
indicate systems with
excesses at both 24~and 70~\micron\ (Table~\ref{excesstable}).
Because the color temperatures
of the excesses are known for these four systems (through
the detection of the excesses at multiple wavelengths),
the locations of these symbols on this plot cannot
change significantly, so no tails are given.}
\label{acrit}
\end{figure}

We apply the Holman \& Wiegert stability criteria to the
debris disks in the 22~systems with excesses
(Table~\ref{excesstable}).
Dust may reside in stable circumstellar or stable
circumbinary regions, or in unstable regions as
defined by Holman \& Wiegert;
these different cases are shown schematically
in Figure~\ref{schematic}.
Twelve of the 22~systems with excess have dust
distances that are much larger than the system's
binary separation, implying a circumbinary debris
disk. Not surprisingly, all but one of these
circumbinary debris disks is in a small
separation system (and the separation for the
one exception is still only 5.3~AU). Seven of the 22~systems
with excesses have dust distances that are much
less than the system's binary separation,
implying circumstellar debris disks.
(We have assumed that the debris disk surrounds
the primary star but we have
no way to verify this
for systems that
are not resolved.)
All of these systems have binary separations
greater than 75~AU.
Three systems with excess do not obviously
have either
circumbinary or circumstellar debris disks,
but rather have dust distances that 
imply unstable locations, that is,
dust distances that are similar to the
binary separations. These three systems (HD~46273, HD~80671, HD~127726) are
discussed further in Section~\ref{interp}.
The dynamical classification of each
system (circumbinary,
circumstellar, or unstable) is listed
in Table~\ref{excesstable}, and a histogram
of systems in these three dynamical states
is shown in Figure~\ref{acrit}.


For the four systems with excesses detected
at both 24~and 70~\micron,
the dust temperature is uniquely fit
to the two-band excess;
these systems are plotted in Figure~\ref{acrit}
with filled symbols. Three of these four systems
have circumstellar dust, and one has 
circumbinary dust.
For systems with excesses at only 
one band,
we show in Figure~\ref{acrit} the
maximum temperature solution and a
range of solutions for each system.
The range of solutions lies
between the maximum temperature solution (symbols)
we derive from the data
and the minimum temperature
(50~K) solution (Section~\ref{mindust}) and its corresponding
dust distances and fractional luminosities (end of
the ``tail'' on the data point).
(Two systems have the reverse situation,
in which the best solution is a minimum temperature
solution and the range of solutions allows warmer
temperatures and consequently smaller dust
distances; these two cases are indicated with
dashed tails in Figure~\ref{acrit}.)
Because for most systems a range of solutions
is allowed, and because the boundaries between
stable and unstable regions may have some flexibility,
we note that some reclassification of dynamical
states may be possible. However, the overall distributions
shown in Figure~\ref{acrit} are unlikely to change
substantially.
For the four systems with excesses detected
at both 24~and 70~\micron,
the dust temperature is uniquely fit
to the two-band excess, and consequently
no range of solutions is shown in Figure~\ref{acrit};
these systems are plotted with filled symbols.

\subsection{Dust and precursor asteroid belts}
\label{belts}

We explain our detections of excess emission conceptually
as due to
a belt
of asteroids that
collide, producing dust.
This dust may be observed at its 
generation location.
For both the circumbinary and circumstellar
cases, the simplest explanation is the presence
of a planetesimal belt near the distance that
we derive.
We ignore the possibility of dust produced from
parent bodies that reside in the unstable region,
as the lifetimes of those
parent bodies in the unstable regions are
prohibitively short \citep{hw99}.
However,
dust can move radially due
to radiation pressure or Poynting-Robertson effect,
and therefore can
have a very different emission temperature
than it had in its generation location
(its asteroid belt).
(In fact, the \citet{hw99} calculations for
stability do not apply for dust, but do apply
to the dust parent bodies.)
This may be
the explanation for dust in unstable regions
in the HD~46273, HD~80671, and HD~127726 systems,
as described in the following section.

\subsection{Dust in unstable regions}
\label{interp}

\begin{figure}
\includegraphics[angle=270,scale=0.60]{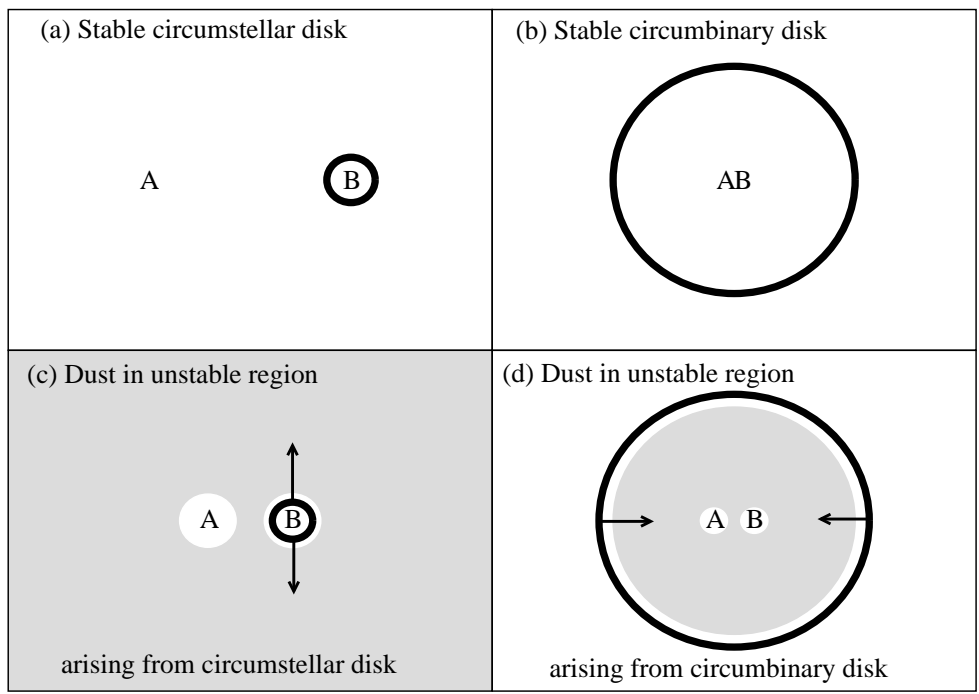}
\caption[]{Schematic diagrams of the four
cases for dust in binary systems.
All features of each panel are to scale.
In all four panels, ``A'' and ``B''
refer to the two stars and the thick, dark
bands represent the location of the
planetesimal population, which is in a stable
location in all four panels.
In all cases, dust is assumed to be produced
in the planetesimal belts.
Cases (a) and (b) correspond to observations
of dust in stable circumstellar and circumbinary
locations, respectively, with the implication
that the dust is observed near its creation location
in the planetesimal belt.
For panels
(c) and (d), the grey shaded area shows
the unstable region as defined in the text.
Cases (c) and (d) show two possible mechanisms
to transfer dust (radial arrows,
ignoring orbital motion) from a stable planetesimal population
to the unstable region where it is detected.
Outward motion, shown in panel (c), is caused
by radiation pressure, while
inward motion, shown in panel (d), is caused
by PR drag.
Case~(d) but not Case~(c) is consistent
with excess emission at 70~\micron\ but
no excess emission at 24~\micron.
All three systems with unstable dust have excesses
only at 70~\micron,
implying that Case~(d) is likely to be the dominant
mode of radial transport.}
\label{schematic}
\end{figure}

Dust in an unstable location could be migrating
inward (under PR drag) from a larger radius (and potentially circumbinary
disk), or could be migrating outward (via radiation
pressure) from
a smaller radius and a circumstellar disk (Figure~\ref{schematic}).
%
%
The parameter $\beta$ is often used to examine
radial migration of dust grains.
It
expresses the ratio
of radiation to gravity forces on an
individual dust grain:
$\beta = (3L_{\star}/16\pi G M_{\star} c) (Q_{pr}/\rho s)$,
where $L_\star$ and $M_\star$ are the stellar luminosity
and mass; $c$ is the speed of light;
$Q_{pr}$ is the radiation pressure coefficient
\citep[we use $Q_{pr}=0.35$, after][]{mm2005};
and
$\rho$ and $s$ are the density and radius of the
spherical grain, with all values in cgs units.
Grains that have $\beta<0.5$ generally spiral inward
under Poynting-Robertson (PR) drag \citep[e.g.,][]{mm2005}.
$\beta$ values
greater than~0.5 correspond to dust particles
that are blown out of the system by
radiation pressure.
A critical size can be derived, where spherical,
solid particles larger
than that size spiral inward under PR drag.
For A5~stars, the critical size is $\sim$5~\micron;
for F5~stars, the critical size is $\sim$2~\micron.
Of course, the picture is somewhat more
complicated when there are two stars
in the system since both radiation and
gravity increase;
furthermore, a particle
may not see a simple radial radiation pattern
as from a single central source. (We revisit
these complications below.)
Particles also need not be solid spheres in 
reality.

We have assumed black body grains, where the particles
are larger than the wavelengths of interest.
Thus, the infrared excesses that we observe
are generally emitted by dust grains larger than
the
critical size of around 5~\micron\ or so.
Such large grains are plausible in debris systems,
although we consider the case of small grains
in Section~\ref{nonbb}.
These grains should therefore be spiraling
inward under PR drag,
suggesting
exterior production zones.

As a further argument,
our temperature (and distance)
derivations solve for the maximum temperature allowed
by the data; in other words, there can be no significant
population of dust interior to the distances that we derive.
Yet, if the dust were migrating outward from an
interior source, there would necessarily
be a population of dust that would be warmer and at smaller
distances than the observed dust location. Dust
closer to the star(s) and warmer would have been detectable
by us, yet
R24 for these two systems is within 3\% of unity.
For these systems,
the dust therefore {\em cannot} originate in 
an interior source region, as we would have detected
it at 24~\micron. 


These arguments are consistent with our
results that the most common configuration
of dust in binary systems is in stable
circumbinary regions. In 12~cases (out of 22~total),
dust is created and observed in circumbinary
locations. In an additional two systems, we suggest,
dust is created in circumbinary locations and migrates
inward under PR drag to an unstable location where 
we observe it (Figure~\ref{schematic}).

\citet{wyatt} and others have suggested that
in debris disks with large enough optical depth
to be observable,
PR drag should {\em never} dominate; instead,
removal of grains that are big enough to be in the PR
(as opposed to radiation pressure) regime will
instead be dominated by collisional processes.
(Note that in the Solar System, whose fractional luminosity
is perhaps a factor of $\sim$100~less than those
for typical
{\it Spitzer}-discovered debris disks, 
PR drag can be a dominant source
of grain removal \citep[e.g.,][]{mm2002}.)
Indeed,
\citet{chen06} provide observational evidence that
IRAS discovered debris disks are dominated by
collisions and not PR drag.
In the planetesimal belt where the dust we observe
is originally produced, it may be true that the
space density of grains is high enough that collisions
dominate. 
We calculate that $\eta_0$, defined by
\citet{wyatt} as the ratio of PR migration timescale
to collisional disruption timescale, is generally
close
to but slightly larger than unity for the disks
detected in this program ($\eta_0\gg1$ means collisions
dominate; $\eta_0\ll1$ means PR dominates). This
implies that 
most grains of the size of interest collide with
other grains before they migrate inward a significant
distance, but also implies that some grains may migrate 
inward under PR drag
without experiencing any collisions.
Furthermore,
the belt where collisions are taking place
must have an inner edge that
is defined by the binary's dynamical interactions.
Grains that leak inward across this boundary may suddenly
be in a region that is devoid of solid material, since
no planetesimals would be in stable orbits there.
In this ``empty'' region, few collisions would take
place and little dust would exist, other than
dust migrating inward under PR drag.
PR effects would dominate since the surface
density of grains would be relatively small, implying
that collisions cannot be a significant loss mechanism.
This may be the dust that we observe in radiative
equilibrium with the binary stars.

\subsection{Detailed analysis of systems with dust in unstable locations}
\label{caveats}

\noindent {\bf HD~46273.} HD~46273 is a member of
a quintuple system.
HD~46273 is the AB pair, with a separation
of 0\farcs 5 or 25.9~AU (Table~\ref{targetinfo}).
The AB-CD separation is 13\farcs 4 \citep{wds}, or about 700~AU (27~times
the AB separation). We detect the CD members of the HD~46273
system in our MIPS images, clearly resolved and outside
our photometric aperture at 24~\micron\ and 
very marginally resolved (and very faint) at 70~\micron. At most,
the CD stars account for 15\% of the measured flux at 70~\micron,
indicating that the excess at this wavelength cannot
be due simply to including flux from the CD stars
in our photometry.
Finally, component A is
a spectroscopic binary, with members Aa and Ab.

Our best fit maximum temperature solution places
the dust at 16~AU.
We recall that the stable regions in a binary system are roughly
bounded by orbital distances
three times less/greater than the binary separation.
Stable orbits are therefore found in the range 80--230~AU
from component A
(bounded on the inside by the AB pair and on the outside
by the AB-CD interaction).
If the Aa/Ab separation is less than 2.6~AU, then
another stable region would exist from three times
the Aa/Ab separation out to 8~AU.
None of these zones clearly imply dust temperatures
of $\sim$100~K. However, 
in light
of the extremely complicated dynamics and radiation
and gravity fields in this system, it is clear
that our first-order calculations of dust
distance and long-term stability may not be
sufficient. 

\vspace{1ex}

\noindent {\bf HD~80671.} HD~80671 is in a triple system, where the AB pair
has a separation of 0\farcs 12, or 3.35~AU
(Table~\ref{targetinfo}). The AB-C separation
is 18\farcs 1 \citep{wds} or 500~AU, around 150~times greater
than the AB separation.
We calculate a dust distance of 2.9~AU.
The existence of component C places no significant
constrains on the location of the dust, and
in this system, it 
is (more) likely that the C component plays
little or no
role in the gravity and radiation
effects of the AB system, where the dust is.
The interpretation may therefore be more straightforward
and more similar to the first-order model we have
presented above.

\vspace{1ex}

\noindent {\bf HD~127726.} HD~127726 is also a triple system, with an
AB separation of 14.3~AU (0\farcs 2).
The AB-C separation is 2\farcs 0, or around 143~AU \citep{wds}.
For this system, the critical outer radius for
the AB pair is nearly equal to the critical inner
radius for the AB-C system, implying that there are
no stable regions in this system between
around 4.5--430~AU. This stability argument leaves
in question where a stable asteroidal population
could reside, from which the observed dust can migrate.
Further study may help reveal
the nature of this interesting system.

We note that all three systems with dust in unstable
locations are multiple (higher than binary) systems.
As argued above, this multiplicity may complicate
the radiation field sufficiently that the dust has a 
temperature that masks its true location in a stable
location. Alternately, more complicated gravitational
interactions may promote asteroid collisions, producing
dust that is short-lived (Section~\ref{residence}).
It is unlikely that a location that is unstable
in a binary system would be stable in a multiple
system, barring unusual resonant configurations.
Finally, for these systems (and others in our
sample) there could be multiple reservoirs of dust,
each having different temperatures and located
in a stable location in the system, that together
masquerade as a dust population with the single
temperature that we derive from a 1~or 2~band excess.

\subsection{Dust at 10--30 AU: Confirmation of expectations}
\label{bias}

Our observations at
24~and 70~\micron\ are generally sensitive to dust
temperatures around 50--150~K,
which are found at distances of 10--30~AU
around A~and F~stars.
Since dust is more likely to be found in
stable regions of binary systems than
in unstable regions (from dynamical arguments),
we are less likely
to detect excess emission from binary systems
with separations of 10--30~AU than from those
with either smaller or larger separations.
Figure~\ref{hist} shows that, indeed, we observe
fewer excesses for intermediate separation
systems,
as we expect.
We conclude that the stability arguments of
\citet{hw99} must apply generally to dust
in binary systems.
We note that this distribution of excess
as a function of binary separation is consistent
with the result of \citet{jensen96}.

The complement of this argument 
is that we are more likely to detect
dust in unstable zones for intermediate
separation systems than for either
small or large separation systems,
for the same reason:
dust at 50--150~K that is unstable is more
likely to be found in intermediate systems (by
the Holman \& Wiegert arguments) than in
either small or large separation systems.
We find this result as well (Figure~\ref{acrit}):
all three systems
with dust in the unstable region are
binaries with intermediate separations,
as expected.

\section{Planetary system formation in binary systems}
\label{planetesimals}

\subsection{Nature versus nurture}

A primary conclusion from this study is that
debris disks exist in binary systems.
Planetesimal
formation clearly is not
inhibited to any great degree by the presence
of a second star in the system.
Whether planetary system formation advances beyond
the planetesimal stage cannot be addressed by 
this work.
However, we can comment on the kinds of
dust-producing interactions taking place among planetesimals
in these systems.

The anti-correlation between age
and debris disks has been well explored
\citep[e.g.,][]{astars,siegler,gorlova,su}, and the
slopes of such functional relationships typically
asymptote to zero for ages 500~million years and
older. For systems older than this, stochastic or
random processes may dominate the production
of dust \citep[e.g.,][]{astars,vegasu}, and not the gradual diminishing of an
initial disk reservoir. However,
a continuous collisional cascade over the lifetime
of the system remains a possible
explanation.


The excess rate we find for binary systems is
marginally higher than that for individual (single)
stars (Section~\ref{context}).
One possibility is that multiple stellar systems may begin
with circumstellar/circumbinary disks that are
more massive than
those surrounding single
stars -- the nature argument.
Hence, multiple stellar systems are
more likely --- from birth --- than individual stars to possess
planetesimal disks and planetary systems,
and therefore
excess emission that we can detect.

Alternately, multiple stellar systems may begin
with circumstellar/circumbinary disks that are
similar to or diminished relative to disks around individual stars.
The multiple stars may then interact in ways
that could cause planetesimal growth (and subsequent
dust production):
the two (or more) stars present in the system
can stir up the circumbinary (or circumstellar)
disk, causing orbits to cross and generally creating
an environment in which accumulation of solid material
in a protoplanetary disk is favored. This is
the nurture argument, that multiple systems
create environments where planetary systems may be
more likely to form \citep{marzari00,boss06}.

We note that this nurture model may have
a significant backlash.
Enhanced dynamical stirring could equally be the
downfall of planet formation if protoplanetary disk
material is excited sufficiently that collisions
are erosive rather than accumulative.
\citet{thebault} 
find that the boundary between these two
regimes depends on the size of the interacting
bodies, among other parameters, implying that 
within a single system collisions could potentially be
erosive for asteroid-sized bodies ($\sim$10~km)
but accumulative for larger bodies.

Determining which of nature and nurture is more
important in planetesimal formation in multiple
systems will require detailed studies of the youngest
multiple systems (e.g., the results presented
in 
\citet{jensen96} and \citet{mccabe06}).
Further disk studies
of an ensemble of
young binary (or higher) systems
will allow us to understand the initial
conditions that lead to the properties of
the sample presented here.

\subsection{Residence times and collisions in binary systems}
\label{residence}

Particles in the unstable zone have
their orbits disrupted in $\lesssim$$10^4$~orbits
of the binary system \citep{hw99}.
The orbital periods for HD~46273, HD~80671, and HD~127726
are all less than 100~years,
so the dust should be removed in less than
1~million years in all cases.
%
The PR crossing
time $\tau_{PR}$
is given approximately by $400 (\Delta r)^2 / M_\star \beta $,
where $\tau_{PR}$ is in years;
$\Delta r$ is the radial distance to be
crossed, in AU; $M_\star$ is in Solar 
masses; and $\beta$ is as defined above \citep{wyatt}.
The unstable zone extends approximately from
0.3~to 3~times the binary separation,
so we set $\Delta r$ equal to 2.7~times the
binary separation.
The PR crossing times, which are the maximum
residence times of dust in the unstable
zone, are therefore also around
1~million years for these systems.
%
These facts together imply that
the dust residing in these unstable
regions is very short-lived
and
that we are witnessing
either the migrating tail of a continuous cascade,
or the result of a very recent collisional event \citep[see, e.g.,][]{lisse07}.
With a few assumptions, we may be able to suggest
which of these is more likely.

For these three infrared excess systems,
the residence time of dust against loss mechanisms
is around $10^{6}$~years and the typical fractional luminosity
is around $10^{-5}$.
For the purposes of this exercise, we assume 10~\micron\
dust grains located 10~AU from a parent star.
In this case, approximately $10^{30}$~grains are required
(at low optical depth) to absorb and re-radiate
the appropriate amount of
emission from the parent star.
For a density of 1~g/cm$^3$, this population of grains
has a mass of $\sim$4$\times10^{21}$~g.
For a residence time of $10^6$~years, this implies
a grain production rate on the order of $10^8$~g/s,
similar to that measured for
comet Hale-Bopp \citep{lisse97}.
This is a small enough production rate that
none of continuous collisional cascade;
stochastic collisions; or individual sources (e.g.,
comets) can be ruled out.
The ages for the two systems with dust in unstable
locations are 1--2~billion years.
Over 1~billion years, the total mass of 10~\micron\
dust grains produced under a continuous collisional cascade would
be around $3\times10^{24}$~g or $5\times10^{-4}$~$M_{\oplus}$,
which implies the efficient disintegration
of $\sim$1000~hundred-km asteroids.
If the 10~\micron\ grains are the tail of a size distribution that
follows a power law that goes as size$^{-3.5}$
\citep[e.g.,][]{dohnanyi}, the total mass
may be 10--100~times greater; a shallower slope \citep[e.g.,][]{reach}
gives an even larger enhancement.
In our Solar System, there are fewer than 400~asteroids
larger than 100~km.
Over the age of the Solar System, the number of 
100~km asteroids in the asteroid belt may
have decreased by only a factor of 5--10 \citep[e.g.,][]{davis}.
Our Solar System's asteroid belt could therefore not
be the source of a continuous dust production
of the magnitude that is presently observed
for these two systems.
The implication may be that the
present dust production rate cannot have been
continuous over the lifetime of the system.
Although other interpretations are possible, this rough
calculation implies that the dust that we observe in
these systems was likely produced in a recent
event, and that
stochastic (occasional) collisions may dominate
the dust production in these systems on billion-year
timescales.

\section{Alternate interpretations and uncertainties}

\subsection{Uncertainties in dust temperatures}
\label{mindust}

Several alternate interpretations can explain
the observations and dynamical stability
constraints.
Recall
that the dust temperatures presented
in Table~\ref{excesstable} and elsewhere are the
maximum temperatures allowed by the multi-wavelength
observations (with the two exceptions described
above and noted in Table~\ref{excesstable}; these
exceptions are addressed below). In general, this is the most rigorous
and most useful statement that can be made, but
cooler temperatures might equally well fit the data
and imply greater orbital distances, with
implications for the implied stability.

To explore the consequences of abandoning our
maximum temperature requirement, we recalculated
dust location and fractional luminosity for 
systems with only 70~\micron\ excesses assuming
that all the dust has a ``minimum reasonable temperature''
of 50~K, a value that is consistent with the smallest excess temperatures
found by \citet{su}.
(Note that this is not truly the absolute minimum
possible, since an undetectable population of very
cold dust cannot be ruled out for any system.)
In all cases 50~K is consistent with a physical
model that can match the observations, and
corresponds roughly to a 24~\micron\ excess
that is $\sim$1\% of the photospheric emission.
Generally, the dust location increases
somewhat in radius
and the fractional luminosity
decreases by up to a factor of~10.
These alternate (minimum) solutions are shown
in Figure~\ref{acrit} as the ends of the ``tails'' extending
from the 
data points. The locus of acceptable solutions
for each system lies along this tail, with the
maximum temperature solution at one end (indicated
by the symbol) and the minimum temperature solution
at the other end.
Alternate solutions, with different temperatures, 
could potentially move systems into or out of the 
unstable zone, and in all cases include a move
to lower fractional luminosities.

For the two systems where the best temperature solutions
are lower limits (HD~118216 and HD~16920), the ``tails''
indicate that warmer temperatures (that is,
smaller dust distances)
are possible solutions.
In each case, assumptions were needed to calculate
the range of acceptable temperatures (Section~\ref{interest});
because of these extra assumptions,
these tails are shown as dashed
lines.

\subsection{Non-black body dust grains}
\label{nonbb}

We have assumed black body dust grains for all
analysis and discussion above, but non-black body
dust grains (generally those with small sizes) 
offer a possible alternative interpretation.
Non-black body grain thermal equilibrium
temperatures are hotter than those of black body-like
grains at the same distance from the star
\citep[see, e.g.,][and Figure~9 therein]{vegasu}.
Therefore, in matching a derived temperature,
non-black body grains 
would imply a
greater distance from the star(s) than the
black body grain solutions discussed above.
Under Mie theory, the dust distance
(in centimeters) is given by
$(R_\star/2) (T_\star/T_{g})^{2.5}$
where $R_\star$ is the stellar radius (in centimeters),
and where the grain
temperatures are those derived in Section~\ref{dustdist}.
Calculating dust distances under this
small-grain assumption moves all systems rightward
in Figure~\ref{acrit}, and moves some systems
from the circumstellar region to the unstable
region, and some from the unstable region to
the circumbinary region. However, the total
number of systems in the unstable region 
is unchanged: even under this small-grain
assumption, there still exist populations
of grains in unstable regions.

Smaller grains might also have $\beta>0.5$.
This means that grains might be spiraling outward
under radiation pressure.
It would still be difficult
to explain any dust that
remains in the unstable zone, since
there is no evidence for hotter
interior reservoirs of dust, as explained above.
A more difficult case would arise if the grain
properties vary significantly from system to
system.
We leave a more complete exploration of these
effects for future work 
as the number of free parameters is large enough
that useful constraints may not be easily obtained.
%

\subsection{Uncertainties in dust location}
\label{unc_dust}

A final uncertainty on our calculations of grain 
temperatures is related to the geometry
of the systems. We have implicitly assumed that
all radiation fields are purely radial, and that
all flux comes from a single source, either as 
a combined close binary or a single star whose
companion is far away. For the two systems where the 
implied dust distances are comparable to the binary
separations --- that is, for systems where the dust
appears to be in the unstable region --- neither
assumption is likely to be true. More sophisticated
modeling that considers the specific geometry of 
a given system would be appropriate in those cases.

There are a number of other potential geometries.
We have assumed in all cases that any circumstellar
dust must be located around the primary, whose spectral
type is known, but of course dust in a binary system
could equally be around either the primary or the secondary.
A more complicated geometry would allow for
circumsecondary dust, but a luminosity ratio for
which the primary significantly heats the dust as well;
such a construction could even masquerade as a dust
population located in the unstable zone.

Following from the previous discussion,
there is potentially substantial uncertainty in the
dust location for any given system.
Even for systems with excesses at both wavelengths,
the possibility of non-black body grains could
allow for a substantial change in dust distance.
There is little to be done to explore the 
consequences of these uncertainties as we have
no more than two excess measurements for any system.
Further observations at different wavelengths, and
especially including low resolution spectra that would
allow us to map the SED at much higher resolution than
broadband photometry permits, are necessary to break
some of these degeneracies and remove uncertainties 
\citep[see, e.g.,][and Figure~13 therein]{su}.
Spatially resolved images at multiple wavelengths would
also help break these degeneracies.

Finally, we model all dust populations as single
populations at a single location. For the four
systems with excesses at both bands, there exists
the possibility that dust could instead reside in
two separate reservoirs: a hot dust population near
one of the stars and a cold(er) circumbinary dust
population. 
The small number of data points does not warrant
further calculations of this possibility, but we 
mention it against the possibility of future
data that may help constrain the location of the dust
in that system.
For the three systems with dust apparently in unstable
locations, having multiple dust populations in 
different (stable) locations could masquerade as
a single population in an unstable location.


\subsection{Uncertainties in binary separations}

In some cases, the separations listed in Table~\ref{targetinfo}
(and Table~\ref{excesstable}) are the
projected separations, not the actual orbital
distance. 
This is because some of these binary systems have not
been monitored long enough or well enough to determine
true orbits for the components. To calculate whether
the dust we observe is in an unstable region, we use
the separations in Table~\ref{targetinfo}, which
lists the best information we have (actual or projected
separations).
Some of our determinations of dust in unstable
regions, therefore, could in theory change categorizations
if the separation information changed substantially.
We suspect that this possibility would contribute only
a minor effect overall as projected separations are not likely
to differ from orbital separations by more than a factor
of~$\sim$2 in most cases.
We also note that higher multiplicity (beyond binarity)
may have an effect in the stability of planetesimal belts
and the production of dust (Section~\ref{caveats}).


\subsection{Possible youth effects}
\label{youth}

Ages are known for most, but not all, of these
69~binary systems (Table~\ref{targetinfo}).
As it is now well established that larger
excesses generally are found around younger
stars \citep[e.g.,][]{astars,su,gorlova,siegler}, 
we
look for possible effects of youth in our results.

\begin{figure}
\includegraphics[angle=270,scale=0.60]{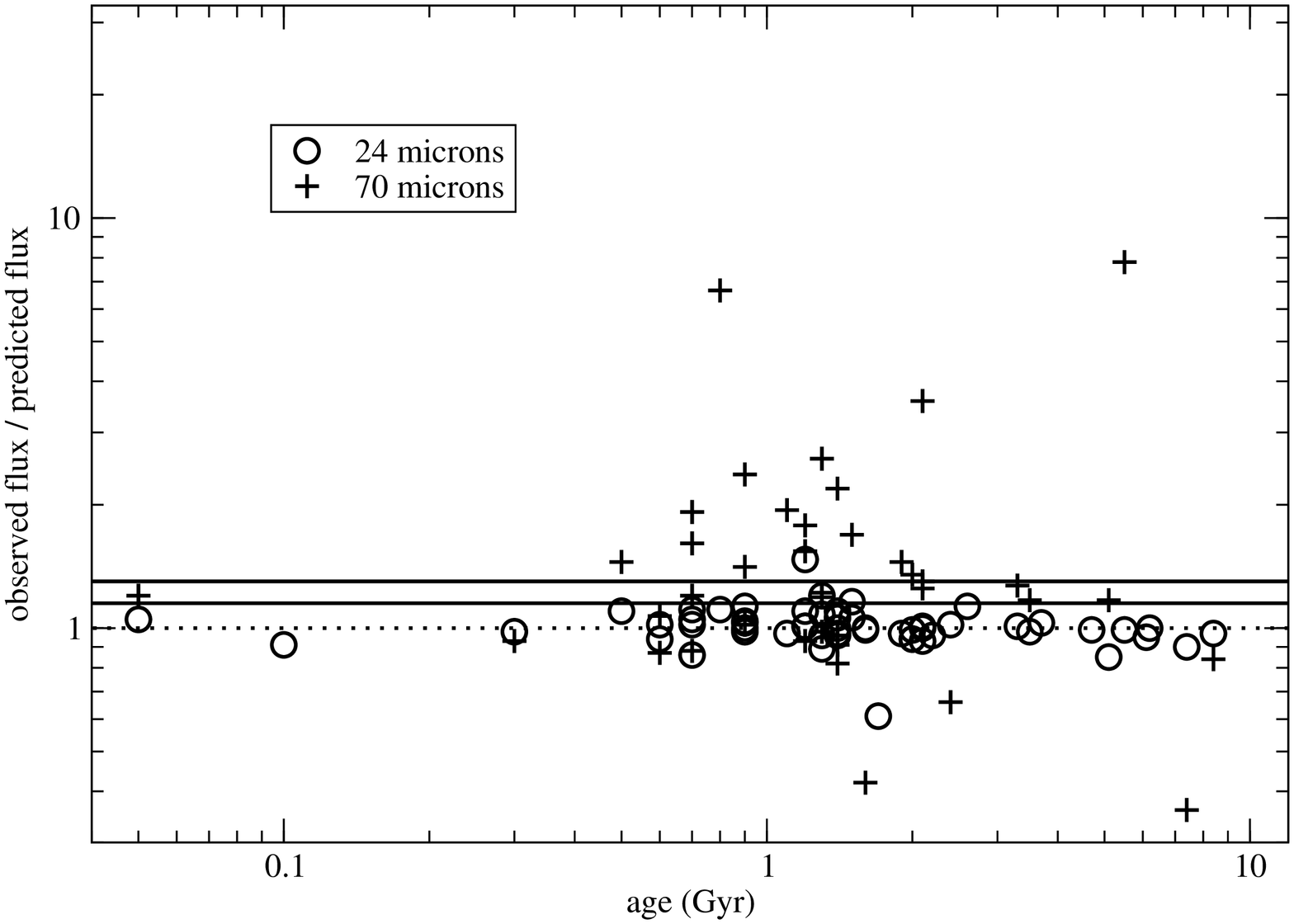}
\caption[]{Ratio of observed to predicted
fluxes for all systems as a function
of system age, for systems with
known ages.
Circles show
data for 24~\micron\ observations and
crosses show results for 70~\micron\ observations.
In all cases the errors in R24 and R70
are smaller than the symbols.
The solid horizontal
lines show our criteria for identification
of excess at
R24~=~1.15 (lower line) and
R70~=~1.30 (upper line), and the
dotted line shows R~=~1.00, for
guidance.
No obvious trend of excess (observed/predicted)
with age is apparent, though we note that
several of the systems with large
excesses have no published ages and hence
are not shown here (see text for discussion);
we cannot rule out the possibility that these systems
with large excesses are young.}
\label{ages}
\end{figure}

R24 and R70
as a function of age
are shown in Figure~\ref{ages} for the 55~systems with
known ages.
One might expect that the younger
systems would be more likely to have detectable
excesses \citep[e.g.,][]{habing,dominik,astars,su,gorlova,siegler}.
However, since most systems with known ages in this program are
older than 1~billion years, and all are older than
around 600~million years (with three exceptions,
as indicated in Table~\ref{targetinfo}),
these systems are all mature, so any age
dependence might be expected to be minor.
Indeed,
we find
no obvious trend of excess with system
age (Figure~\ref{ages}),
though we note that
several of the systems with large
excesses have no published ages and hence
are not shown here (see text for discussion);
we cannot rule out the possibility that these systems
with large excesses are young.

There are 14~systems for which 
no published or calculated ages are available.
However, we can classify ten of these systems
as ``old'' or ``young,'' as follows; since
we are looking for possible effects of youth,
this rough classification suffices.

HD~83808, HD~13161 (a large excess system), HD~17094, and HD~213235 are all more luminous
than dwarf spectral classes, according to the
Gray \& Corbally technique (see Table~\ref{targetinfo} and
Appendix~A).
These systems (or at least the primary stars) are therefore
likely at the old end of their main sequence lifetimes
(and have started to evolve off the main sequence).
HD~137909 is part of an extensive study by \citet{hubrig},
who put it on the HR diagram
and find it to be well above the ZAMS,
implying that this system is old.
These systems are labeled ``old'' in Table~\ref{targetinfo},
indicating that excesses around HD~83808, HD~13161, and HD~17094
cannot be due to youth effects.

HD~127726 (70~\micron\ excess) and HD~50635 (no excesses)
were both detected by ROSAT \citep{hunsch}.
Both stars have spectral types F0V, which may be too early
for chromospheric activity, so a likely scenario
is the presence of an active late-type dwarf companion.
This argues for youth for these systems, but is not a
strong constraint, since the main sequence lifetimes of
late-type dwarfs are much longer than those for F0 main sequence
stars.
We label both HD~127726 and HD~50635
``young'' in Table~\ref{targetinfo}.
Indeed,
\citet{barry88}
estimate a chromospheric age of 300~Myr for HD~50635
based on Ca II H and K measurements of
the secondary.
%
However, Str\"{o}mgren photometry \citep{mermilliod} of the HD~50635 primary alone is
available (due to the system's relatively
wide separation), and suggests an age of 990~Myr (using the \citet{moon85}
grid of stellar parameters), indicating that our designation
of ``young'' for this system has substantial uncertainty.

HD~61497, like HD~50635, is widely enough separated
that the primary alone can be identified in Str\"{o}mgren photometry \citep{mermilliod}.
Again using the \citet{moon85} grid of stellar parameters,
we estimate an age of 520~Myr for this system, and we
therefore label it ``young'' (following our definitions
of ``young'' and ``old'' in Section~\ref{context}).

Finally,
we use the $\log g$ values presented in Table~\ref{targetinfo}
to conclude that HD~72462 is old and that HD~6767 is young,
based on their low and high gravities, respectively.
HD~61497 also apparently has a low gravity, but because of the
poor fit in the Gray \& Corbally technique we defer
to the Str\"{o}mgren photometry technique described above.

Four systems remain with no age estimates.
HD~95698 (the system with the largest excess
at 70~\micron, with R70~=~25.56) and HD~77190 have 
$\log g$ between 3.8~and~4, which is inconclusive
for determining ages.
We have no $\log g$ measurements for
HD~16628 and HD~173608.

Of the four ``young'' systems and three systems
identified as young by I.\ Song (pers.\ comm.), only HD~127726 has
an excess. Of the four systems with no age estimates,
three have excesses, including
HD~95698, whose R70 value is reminiscent
of a young system \citep[e.g.,][]{su}. Results from these subsamples
are obviously hampered by small number statistics, 
but the excess rates are not too different from those
measured by \citet{su} for young A~stars, but
are also not too different from the overall excess
rates for this binary sample. Furthermore, removing
these eight systems (four ``young'' plus four
with no age estimates) from our binaries sample does
not significantly change the observed frequency
of disks at either 24~or 70~\micron.
We therefore conclude that there is no 
overall bias due to young systems in our results.

\section{Comparison to other debris disk results}
\label{comparison}

\subsection{Context: Debris disks in non-binary systems}
\label{context}


\citet{astars,serena,bryden,su,siegler,gorlova,tpf}
and others have recently used {\it Spitzer} to study
the fraction of
AFGK stars with debris disks; most of the targets
in those samples are in non-multiple systems.
Disentangling age effects from spectral
type is difficult, but
our excess fraction results can be roughly
placed in context as follows.

A primary conclusion of many of those previous studies
is that age is the dominant factor in determining 
excess fraction (the number of systems with excesses):
the excess fraction decreases with increasing age.
Therefore, to place our binary debris disk results
in context with results from non-binary systems, we need
to compare to populations of similar age.
Since all but three of the systems with known ages
in our sample are older than 600~Myr, we use that 
criterion here.

The relevant comparisons are to the overall
excess rates for ``old'' A~and F~stars (the latter of which we summarize
here as FGK~stars as there are no published
results for a large sample of just F~stars),
where 
``old'' is defined as $>$600~Myr.
At 24~\micron,
the excess rates for old A~and FGK~stars
are around 7\% and 1\%, respectively.
After discarding the 11~systems (see above) in our
sample that either have ages less than 600~Myr (three 
systems); are ``young'' (four systems); or have no age
estimates (four systems), our excess rate
is 9$^{+5}_{-3}$\%, marginally 
higher than the results for single
AFGK stars (and perhaps significantly higher
than the results for FGK stars).

At 70~\micron,
the excess rates for old A~and FGK~stars
are 25\% and 15\%.
Our excess rate for the 42~old
systems that were observed at 70~\micron\
is 38$^{+8}_{-7}$7\%, again 
marginally higher than the rate observed
for single AFGK stars (and again perhaps significantly
higher than the results for FGK stars).

%

Excess rates that
are marginally higher than those
for individual (non-multiple) AFGK stars
may argue that binary
systems are more likely
to have planetesimal
belts than single stars. Alternately,
it may argue that 
planetesimal belts in binary systems
are similar to those of single stars,
but
more likely
to be in an excited (i.e., recently
collided) state.
This is the nature/nurture argument
about binary systems and planetesimal
formation presented above.

Around 45\% of wide binary systems have
debris disks (Figure~\ref{hist}). The
disks in these systems are generally
very close to the primary (assumed) and 
far from the secondary.
It might therefore be argued that the 
secondary has little to
do with the presence or absence of disks
in widely separated binaries, and with
our data we cannot eliminate the possibilities
that disks exist around the secondaries in
these systems, either instead of or in
addition to disks around the primary.
The excess rate might therefore be given
as 20\%--25\% per star for widely separated
binary systems.
This number is quite consistent with the
excess rates measured in the surveys
of single stars listed above.

We make the above point about wide binaries in
order to emphasize the fact that the excess
rate for small separation binary systems
is nearly 60\%. These debris disks are 
circumbinary, so it is clear that the presence
of the secondary star cannot be ignored when
considering the evolution of the debris disk.
Small separation binary systems must clearly,
in some way, promote the presence of the kind
of debris disk that we can detect.
It would be an interesting further observational
and theoretical study to understand why the
detectable debris disk rate for small separation
binaries is so much larger than that for single
stars (or for wide separation binaries per star).

Debris disks around A~stars typically have fractional
luminosities around $10^{-4}$ to $10^{-5}$,
with only protoplanetary or intermediate-age disks
substantially larger \citep{su}.
The fractional luminosities of disks around old FGK
stars are typically a few times 10$^{-5}$.
The typical fractional luminosities we report
here are similar to the results
for those two samples, as expected.




\subsection{Debris disks in binary systems in other samples}

%
%
%
We look for binary systems in the 
\citet{bryden}, \citet{tpf}, and \citet{su} samples
to extend our results.
By design, there are few binary
systems in these samples, and there are
only 8~binary systems total
that are ``old'' ($>$600~Myr).
Of these~8, only one, HD~33254, has an excess
at 70~\micron, and none have excesses at
24~\micron.
If we aggregate these 8~systems with our
sample, the excess rates at 24~and 70~\micron\
go down slightly, to 9\% and 36\%, respectively.
This result is still marginally high compared
to the excess rates for old single AFGK stars;
because of the 
small number of additional targets and
detections, data on these 8~systems add little
to our understanding of planetary system formation
in binary systems.

\citet{chas2}, in a preliminary result, found that 6~of
26~FGK stars with known extrasolar planets (23\%) 
show excess emission at 70~\micron\ (none
of the 26~have excesses at 24~\micron),
a result that would be marginally different from
field FGK stars without known planets. (Stars with
known extrasolar planets are generally ``old'';
the vast
majority of
known extrasolar planets are on orbits
comparable to or smaller than the binary orbits
presented here.)
However, \citet{bryden} and Trilling et al.\ (in prep.),
extending the work of \citet{chas2},
found that 
the excess rate enhancement for FGK stars
with known planets is marginal at best.
Nevertheless, 
we note that both small separation binary
systems (from this work) and planet-bearing systems
\citep{chas2}
may be (more) likely to have debris disks
(though with a fair amount of uncertainty
for the results for both populations).
The mechanism(s) for planet formation
may be very different from those
of binary star formation,
but
broadly speaking both are 
binary (or multiple) systems, perhaps
suggesting a commonality of properties.
Again, further observations will be necessary
to probe this possible connection.


\section{Conclusions and implications for planet formation}
\label{conclusions}


We observed 69 main sequence A3-F8 binary 
star systems at 24~\micron, and a subset of 53~systems
at 70~\micron, to look
for excess emission that could suggest 
dust grains and, ultimately, planetesimals
and planetary systems.
We detected excess emission (observed fluxes
greater than predicted photospheric emission by at 
least 3$\sigma$) 
from 9$^{+4}_{-3}$\% of our sample at 24~\micron\
and 40$^{+7}_{-6}$\% of our sample at 70~\micron.
Four systems show significant excess at 
both wavelengths.
We interpret this excess emission as arising
from dust grains in the binary systems, leading
to our first main conclusion: {\em binary systems
have debris disks}. The incidence of debris 
disks is around 50\% for binary systems
with small ($<$3~AU) and with large ($>$50~AU) separations.

For systems with excess emission, we compute or 
constrain the dust temperature, assuming blackbody
emission, and use that temperature to model the location
of the dust within the system.
Dynamical stability arguments suggest that
in more than half of the cases with detected
excesses the dust is probably circumbinary;
that
another third of the sample of systems with
known excesses have circumstellar dust;
and that, in our second main conclusion,
{\em three multiple systems
have dust in dynamically unstable locations.}
This situation likely arises when 
dust produced by collisions in a circumbinary
disk migrates inward under Poynting-Robertson (PR)
drag to its presently observed location. 
The dust residence times in the unstable regions
are less than 1~million years against both
dynamical effects and Poynting-Robertson drag,
implying that we are witnessing
either the migrating tail of a continuous cascade,
or the result of a recent collisional event.
For two of the three systems with dust in
unstable locations, there is no obvious location
in the system where a stable reservoir of precursor
asteroids could reside.


Our third main conclusion is that
{\em the overall excess rate that we measure
is marginally higher than that of individual
(single) old ($>$600~Myr) AFGK stars.}
Binary star systems are therefore at least as
likely as single stars to possess
debris disks and, by implication, planetesimal populations.
Planetesimal formation
may have proceeded because
of the nature of the system if the processes
that form multiple stellar systems also produce
environments conducive to the accumulation of 
small bodies.
Alternately,
planetesimal and planetary system
formation may take place
in multiple star systems if
the stars themselves create a dynamical environment that
promotes accumulation of planetesimals -- the nurture
argument.
A twist on the latter model is that extreme
dynamical stirring in multiple systems may
delay or preclude the formation of big bodies
when collisions become erosive.

Our fourth main conclusion is that {\em the
excess rate (fraction of systems with
excesses) for small separation ($<$3~AU)
binaries is quite high, at nearly 60\%}.
Either
the processes that form small separation
binaries must also commonly form debris
disks (the nature argument),
or else small separation binaries evolve
in a way
that is
likely to produce debris disks that
we can observe (the nurture argument).

Around 30\% of known extrasolar planets are
found in widely separated binary systems, and the
recent discovery \citep{konacki} of an extrasolar planet
in a tight triple system further indicates that
stellar multiplicity does not
preclude planet formation.
We have found that dust production and the implied
presence of planetesimals is at least as common
in binary systems as it is for individual stars.
Determining which of nature and nurture is more
important in planetesimal formation in multiple
systems requires detailed studies of the youngest
multiple systems to track the creation of
planetesimal belts.



\acknowledgments

We thank
Norm Murray for pointing us to the 
Holman \& Wiegert paper and
Nick Siegler and Morten Andersen for interesting discussions.
We thank Steve Kortenkamp for thoughts on the evolution
of the asteroid belt and Inseok Song for sharing his
results on the ages of several systems.
We acknowledge useful suggestions made by
an anonymous referee.
This work is based in part on observations made with the {\it Spitzer} Space
Telescope,
which is operated by the Jet Propulsion Laboratory, California Institute of
Technology
under NASA contract~1407.
Support for this work was provided by NASA
through
Contract Number 1255094 issued by JPL/Caltech.
This publication makes use of data products from the Two Micron All Sky Survey, which is a joint project of the University of Massachusetts and the Infrared Processing and Analysis Center/California Institute of Technology, funded by the National Aeronautics and Space Administration and the National Science Foundation.
The Heinrich Hertz Submillimeter Telescope (SMT) is operated by the Arizona Radio Observatory (ARO), Steward Observatory, University of Arizona.
This research has made use of the SIMBAD and
Catalogues databases
operated at CDS, Strasbourg, France, and of
NASA's Astrophysics Data System.



{\it Facilities:} \facility{Spitzer(MIPS)}, \facility{HHT}.

\appendix

\section{Physical parameters of sample members}
\label{details}

Here we describe our various techniques for deriving
the stellar properties
that
we report in Table~\ref{targetinfo}.
In all cases, we calculate stellar temperature through two
independent methods, described here, to ensure the quality
of
our photospheric fits.

\subsection{Kurucz models}

We use visible and near-infrared photometry
from the literature (primarily Hipparcos and 2MASS)
to derive a best-fit photospheric model. In many
cases, the binary components are resolved in these
literature catalogs, but in almost all cases are
not resolved in our {\em Spitzer} images. We therefore
combine the fluxes at every wavelength to produce
system-integrated fluxes, which are then used in
our photospheric modeling.
Five systems are resolved at 24~\micron\
but not 70~\micron; in these cases, the 
system-integrated approach is used by
summing the fluxes for the two components
at 24~\micron. Five additional systems
are resolved at all wavelengths; for these 
systems only, the photometry for only
the primary is used.
Additional details of this photometry
are given in Table~\ref{photom} and Section~\ref{spitzer}.

We derive a best-fit Kurucz model 
by fitting all available optical to near infrared
photometry (Johnson $UBVRIJHK$ photometry [JP11, \citet{jp11}],
Str\"{o}mgren $uvby$ photometry \citep{uvby},
Hipparcos Tycho $BV$ photometry \citep{tycho}, 
2MASS $JHK_s$ photometry \citep{2mass}) based on a
$\chi^2$ goodness of fit test.
We employ a grid of Kurucz models in our search for best fits, with
effective temperature step size 250~K and metallicity
unevenly spaced between +1.0~and~-5.0 \citep{kurucz93,castelli03}. 
Extinction is a free parameter in our photosphere fits,
and for 26~systems is non-zero (see Table~\ref{targetinfo}).
We estimate the extinction ($A_V$) based
on the $B-V$ color and spectral type, and then apply dereddening based
on the extinction curve from \citet{cardelli89}.
The resulting best-fit parameters, reported
in Table~\ref{targetinfo}, are effective temperature
and metallicity (which determine the best Kurucz model)
and extinction.


\subsection{Spectroscopic analysis (Gray \& Corbally)}

The technique used to derive the Gray \& Corbally
parameters presented
in Table~\ref{targetinfo} is identical to that reported
in \citet{gray03} and \citet{gray06};
here we briefly summarize that technique.
Classification-resolution spectra (1.8--3.5~\AA) were
obtained with the Dark Sky Observatory (DSO) 0.8-meter telescope of Appalachian
State University, the CTIO 1.5-meter telescope, and the 2.3-meter Bok
telescope of Steward Observatory (SO).  The DSO spectra have a spectral
range of 3800--4600~\AA, with a resolution of 1.8~\AA; the CTIO spectra a
spectral range of 3800--5150~\AA, with a resolution of 3.5~\AA; and
the SO spectra a spectral range of 3800--4960~\AA\ with a resolution of
2.6~\AA.
These spectra were rectified and classified by direct
visual comparison to MK standard stars.

The physical
parameters of the stars are determined by simultaneous fitting between
the observed spectrum and a library of synthetic spectra; and between
observed mediumband fluxes (from Str\"{o}mgren photometry) and theoretical
fluxes based on model atmospheres (there is no available Str\"{o}mgren
photometry for HD 16628).
Most systems required zero reddening in this technique,
except for HD~199532 (where $E(b-y)=0.061)$) and HD~8224 ($E(b-y)=0.04$)
(where $E(B-V)=E(b-y)/0.74$).
Chromospheric activity is also identified (see below),
where present.
Further details of all of these steps are explained
in the two papers referenced above.

A number of these stars appear
slightly peculiar and mildly metal-weak, likely due to the
composite nature of these spectra.
In the spectral types in Table~\ref{targetinfo}
the ``k'' type
refers to the K-line and the ``m''
type to the general strength of the metallic-line
spectrum.
The first part of the spectral type generally -- with the
exception described below -- is most strongly correlated
with effective temperature.
Therefore, HD~151613 has a spectral
type
of F4~V~kF2mF2, and so has an effective temperature close to that
of a normal F4~star and a K-line and a metallic-line
spectra that are similar to those of an F2 star.
The exception to the effective temperature
guideline is the Am stars.  For instance,
HD~207098 has a spectral type
kA5hF0mF2~III. For this ``h'' or hydrogen spectral type,
F0 will be the best indicator of the effective temperature.
The difference in the notations is because Am stars generally appear
metal-rich. 

\subsection{Simbad information}

For cases in which we do not have effective temperatures
from the Gray \& Corbally technique, we calculate effective
temperatures based on the Simbad spectral type and interpolations
of the relationship between spectral type and effective
temperature given in \citet{drilling}.
These values are given in Table~\ref{targetinfo}
in parentheses in the ``Teff (G\&C)'' column.

\subsection{Comparisons among derived types and temperatures}

Each of these three fitting techniques was carried
out independently of the others.
Generally the match between the Simbad spectral types
and the G\&C types is quite good. The only system where
there is a substantial mismatch is HD~83808,
where Simbad gives A5V but G\&C derive
F6III, but this disagreement is consistent with
the analysis of \citet{hummel}, who found that the primary is F9III (e.g., G\&C) and
the secondary is A5V (e.g., Simbad).
Since this system harbors excesses
at 24~and at 70~\micron, further
study would be appropriate to constrain the 
excesses more precisely. G\&C note that
their fit produces large photometric residuals, and that
the 
spectrum is clearly composite with a weak Ca K line.

The best-fit temperatures we derive from the Kurucz
models generally agree quite well with the G\&C derived
temperatures; only eight system effective temperatures
differ by more than 500~K, of which two G\&C fits
have large photometric errors.
Temperature errors of 500~K result in predicted
24~and 70~\micron\ fluxes that change by less
than 1\%, which is significantly less than our
calibration errors in all cases, so we conclude
that temperature fits that are good to within
500~K are sufficient for our needs.
The metallicity matches are slightly less good, with a
number of
systems having significantly different metallicities
in the two derivations (our Kurucz fits and the
G\&C derivations).
This is not overly surprising, since the broadband
photometry approach we use to find the best-fit Kurucz
models does not constrain metallicity well and is
not very sensitive to the values given in
Column~4 of Table~\ref{targetinfo}.
Nevertheless, the overall agreement
between the two techniques is quite satisfactory,
and serves as another positive cross-check that our
photospheric fitting technique is adequate
for the work presented here.
All of the physical properties discussed here are
presented in Table~\ref{targetinfo}.

\section{Notes on individual sources}
\label{notes}

%

\noindent {\bf HD~95698.}
This system has the largest excess ratio
in our sample, with R70 of more than~25.
Formally there is no excess at 24~\micron,
but R24 is close to and $\chi_{24}$ is greater than
their threshold values, perhaps implying
a small excess at this wavelength (see below).
Although the 70~\micron\ excess is quite large, the fractional
luminosity is only slightly high compared to the rest 
of our sample because the 
maximum color temperature of the excess is relatively
cool (65~K). Therefore the
thermally emitting dust
is at a large distance
from the binary star system.

R70 of~25 is quite large.
Only two ``old'' A~stars (Section~\ref{context}) in the \citet{su}
sample have excesses close to this large, and no
``old'' stars in the \citet{bryden}
FGK sample do.
However, this excess ratio would be
somewhat unremarkable for a young star.
Since the age of HD~95698 is not known,
it may be an
intriguing candidate for follow-up observations,
but it may also be found that this system is relatively
young and therefore not that unusual.

\vspace{1ex}

\noindent {\bf HD~119124.}
This system shows the largest 70~\micron\
excess of any in our sample with
no indication of 24~\micron\ excess.
\citet{chen05} observed this system independently
with MIPS
and also identified this system
as having an excess at 70~\micron; intriguingly,
they also estimate an age for this
system of $<$200~Myr through its possible
association with the Castor moving
group, quite in conflict
with the \citet{nord04} age of 5.5~Gyr.
The large 70~\micron\ excess
for this system may indeed imply that the young
age estimate is more appropriate.
Our measured 70~\micron\ flux is somewhat higher
than that of Chen et al.\ (our 73.65$\pm$6.83~mJy compared to
their 56.1~mJy), though they probably agree at a $\sim$2$\sigma$~level.
Our re-reduction of their (independent) data (from the {\it Spitzer} archive)
gives 24~and 70~\micron\ values quite close to our
values in Table~\ref{photom}, so the discrepancy
may have its root in differences in data reduction
and photometric calibration (the updated calibration factor
we use contributes a small factor (5\%)
in the correct direction).
Our calculated dust temperature of
81~K is higher than their 40~K, and
consequently our fractional luminosity of
$6\times10^{-5}$, under
the maximum temperature assumption, is higher
than their estimate of $2.6\times10^{-5}$.

\vspace{1ex}

\noindent {\bf HD~99211.}
HD~99211 was identified as Vega-like by \citet{mannings},
meaning it possesses an infrared
excess that is likely attributable to
a dusty debris disk. We also find excess emission
from this system at 70~but not 24~\micron.

\vspace{1ex}

\noindent {\bf HD~95698, HD~20320, HD~217792, HD~88215.}
These four systems all have formal excesses at
70~\micron.
Formally, none of them have excesses at 24~\micron,
but in all cases R24 nearly exceeds the threshold
value of~1.15 and $\chi_{24}$
exceeds the significance threshold of~2.0.
Consequently, for these four systems, it is likely
that we
detect excess emission at 24~\micron, though
not at a significant level, and we consequently
treat these systems as having single band excess detections.
For these systems,
we compute dust temperatures and related quantities
as described above, using the measured 70~\micron\
flux and the predicted 24~\micron\ photosphere,
plus three times the relevant error. In all cases,
this prediction+3$\sigma$ value is consistent with
the observed 24~\micron\ flux.

\vspace{1ex}

\noindent {\bf HD~26690, HD~31925, HD~200499, HD~51733.}
HD~26690 has R24 below but near
the threshold; $\chi_{24}$ above the threshold;
R70 near unity; and $\chi_{70}$ near zero.
This system could have an excess that is seen
only at 24~\micron, making it similar to HD~16920,
above.
HD~31925 has R70 just below the threshold
and $\chi_{70}$ above the threshold and no signs
of excess emission at 24~\micron. This system therefore
may have a small 70~\micron\ excess that is not formally
detected by us.
HD~200499 and HD~51733 both have R24 below but near
the threshold; $\chi_{24}$ above the threshold;
R70 above the threshold; and $\chi_{70}$ below
the threshold.
The interpretation of this excess pattern
is unclear, but may suggest a weak excess at
both bands.

\vspace{1ex}

\noindent {\bf HD~99028, HD~29140, HD~80671, and other triple systems.}
HD~99028 is a triple system, with inner orbital
distance of 1.9~AU \citep{soder99} and outer orbital radius $\sim$20~AU
\citep{roberts05}. As discussed in Section~\ref{unstable}, these two orbits
allow essentially no intermediate dynamically stable
regions, and the only stable zone in this system would be
far outside the outer orbit.
It may not be surprising, in
this case, that no excesses are observed in this system.
We list the outer orbit in Table~\ref{targetinfo}.

The
SB9 catalog \citep{sb9} lists two periods for HD~29140,
1350~d and 3.6~d (corresponding to
orbital distances $\sim$0.04~AU and
$\sim$3~AU), likely
indicating that this system is triple.
The intermediate stability zone in this system
must also be quite small, though orbits outside
of $\sim$10~AU may be stable. We list both orbits
in
Table~\ref{targetinfo}, but the outer orbit
may be more important in determining stable regions
in that system.

HD~80671 is also a triple system, and a system
with an excess in an unstable location. This is
system is discussed in detail in Section~\ref{caveats}.

The complications arising from studying
dust in binary systems clearly are magnified for
triple systems (Section~\ref{caveats}).
A number
of other systems in our sample likely
have additional system members (known or
unknown).
The regions of dynamical stability --- if any ---
in triple and higher systems are no doubt
much more complicated than in binary systems.
It may be that excess emission is less likely for
higher multiplicity
due to increased dynamical interactions.
However, we note that all three systems with
dust in unstable locations are found in
multiple (triple or higher) systems: excesses
and dust are clearly not prohibited.





\clearpage

\begin{deluxetable}{lc|ccc|ccccc|ccc}
\tabletypesize{\scriptsize}
\rotate
\tablecaption{Target information\label{targetinfo}}
\tablewidth{0pt}
\tablehead{
\colhead{Name} 
& \colhead{Spec.\ type} 
& \colhead{Teff} 
& \colhead{[M/H]} 
& \colhead{$A_V$} 
& \colhead{Spec.\ type\tablenotemark{c}} 
& \colhead{Teff} 
& \colhead{$\log(g)$} 
& \colhead{v$_t$} 
& \colhead{[M/H]} 
& \colhead{Sep.}
& \colhead{Sep.} 
& \colhead{Age} \\
\colhead{}     
& \colhead{[Simbad]}         
& \colhead{[Kurucz\tablenotemark{a}]} 
& \colhead{[Kurucz\tablenotemark{a}]} 
& \colhead{fit\tablenotemark{b}} 
& \colhead{G\&C} 
& \colhead{G\&C} 
& \colhead{G\&C} 
& \colhead{G\&C} 
& \colhead{G\&C} 
& \colhead {}     
& \colhead{}  
& \colhead{} \\     
\colhead{}  
&\colhead{}  
& \colhead{(K)} 
& \colhead{} 
& \colhead{} 
& \colhead{} 
& \colhead{(K)} 
& \colhead{[cm/s$^2$]} 
& \colhead{km/s} 
& \colhead{} 
&\colhead{(AU)}  
&\colhead{(arcsec)}  
&\colhead{(Gyr)}} 
\startdata
HD 207098 & A5m  & 8250 & -3.0 & 0.3 & kA5hF0mF2 III   &    7301 &    3.66 &     2.0 &   -0.13 & 0.025\tablenotemark{j} & $<$0.01 & 0.6\tablenotemark{y} \\
HD 106112 & A5m & 7250 & +0.3 & 0.0 & kA6hF0mF0 (III)    &  7249  & 3.83 &  2.0  &  0.36 & 0.03\tablenotemark{k} & $<$0.01 & 0.9\tablenotemark{y} \\
HD 118216 & F2IV & 7250 & -0.3 & 0.5 & F3 V comp\tablenotemark{f} & 6582 & 3.10 & 2.5 & -0.32 &  0.04\tablenotemark{n} & $<$0.01 & 1.2 \\
HD 150682 & F2IV & 6750 & -0.5 & 0.1 &  & (6750) & & & & 0.04\tablenotemark{o} & $<$0.01 & 2.1 \\
HD 29140 & A5m & 8250 & -0.1 & 0.15 & A5 IV & 7837 &  3.85 &  2.3 &  -0.08 & 0.04\tablenotemark{l}, 3.0\tablenotemark{m} & $<$0.01, 0.07 & 0.7\tablenotemark{y} \\
HD 199532 & F4IV & 7750 & -3.0 & 0.9 & F5 III-IV & 6571 & 3.60 & 2.4 & -0.03 &  0.05\tablenotemark{p} & $<$0.01 & 0.9 \\
HD 119756 & F3V & 7000 & -2.0 & 0.0 & F2 V               &    6781 &    3.98 &     2.0 &   -0.09 & 0.1\tablenotemark{q} & $<$0.01 & 1.6 \\
HD 16920 &  F4IV & 7000 & -0.1 & 0.0 &   F5 V Fe-0.5 & 6549 & 3.78 & 2.0 & -0.32 &  0.12\tablenotemark{q} & $<$0.10 & 1.3 \\
HD 20320 & A5m & 7750 & +0.2 & 0.0 & kA4hA9mA9 V        &    7680 &    3.94 &     2.0 &    0.04 & 0.17\tablenotemark{q} & $<$0.01 & 0.8\tablenotemark{y} \\
HD 204188 & A8m & 7750 & -3.0 & 0.0 & & (7750) & & & & 0.18\tablenotemark{r} & $<$0.01 & 0.1\tablenotemark{y} \\
HD 83808\tablenotemark{d} & A5V & 6750 & +0.5 & 0.0 & F6 III Sr Ca wk\tablenotemark{f} & 6570\tablenotemark{i} & 2.88\tablenotemark{i} & 4.0\tablenotemark{i} & -0.16\tablenotemark{i} &  0.19\tablenotemark{s} & $<$0.01 & old \\
HD 88215 & F2 & 7000 & +0.1 & 0.0 & F2 V               &    6776 &    3.96 &     2.0 &   -0.19 & 0.2\tablenotemark{t} & $<$0.01 & 0.7 \\
HD 178449 & F0V & 6750 & +0.0 & 0.1 &  & (7000) & & & & 0.3\tablenotemark{q} & $<$0.01 & 1.4 \\
HD 13161 & A5IV & 8500 & -3.0 & 0.0 & A5 IV              &    8186 &    3.70 &     2.0 &    0.20 & 0.3\tablenotemark{u} & $<$0.01 & old \\
HD 217792 & A9V & 7500 & -3.0 & 0.0 & F1 V Fe-0.8        &    7143 &    3.96 &     2.0 &   -0.30 & 0.7\tablenotemark{q} & 0.02 & 0.9\tablenotemark{y} \\
HD 11636 & A5V & 8500 & -1.5 & 0.0 & kA4hA5mA5 Va       &    8300 &    4.10 &     3.5 &    0.02 & 0.66\tablenotemark{u} & 0.04 & 0.3\tablenotemark{y} \\
HD 151613 & F2V & 6750 & -0.3 & 0.0 & F4 V  kF2mF2 & 6669 & 3.90 & 2.0 & -0.32 &  1.14\tablenotemark{q} & 0.04 & 1.9 \\
HD 99028 & F2IV & 7000 & +0.1 & 0.2 & F5 IV              &    6600 &    3.70 &     2.7 &   -0.03 & 1.91\tablenotemark{v}, 20\tablenotemark{w} & 0.08, 0.8 & 1.2 \\
HD 17094\tablenotemark{e} & F1IV & 7500 & -0.5 & 0.2 &  A9 IIIp            &    7225 &    3.90 &     3.2 &    0.04 & 2.58 & 0.10 & old \\
HD 32537 & F0V & 7250 & -0.2 & 0.2 & F2 V               &    7018 &    4.05 &     2.1 &   -0.12 & 2.62 & 0.10 & 1.4 \\
HD 17206 & F5 & 7000 & +0.1 & 0.3 & F6 V               &    6378 &    4.06 &     2.0 &   -0.02 & 2.80 & 0.20 & 3.5 \\ \hline
HD 80671 & F4V & 7000 & +0.3 & 0.0 & F5 V Fe-0.7 CH-0.5 &    6618 &    4.05 &     1.8 &   -0.31 & 3.35 & 0.10 & 2.1 \\
HD 56986 & F0IV & 7000 & -0.1 & 0.0 & F2 V kF0mF0        &    6906 &    3.68 &     2.6 &   -0.27 & 3.60 & 0.20 & 1.3 \\
HD 26690 & F2V+F5V & 7000 & +0.0 & 0.0 & F2 V               &    6820 &    3.92 &     2.0 &   -0.15 & 3.70 & 0.10 & 1.4 \\
HD 10009 & F7V & 6250 & +0.1 & 0.0 &  F8.5 V Fe-0.5      &    6162 &    4.14 &     1.0 &   -0.18 & 3.82 & 0.10 & 4.7 \\
HD 195725 & A7IV & 8250 & +0.3 & 0.3 & & (8000) & & & &  4.16 & 0.10 & 0.7\tablenotemark{y} \\
HD 95698 & F1V & 7250 & -0.5 & 0.0 & F2 V & 7042 & 3.80 & 2.4 & -0.11 & 5.31 & 0.10 & \nodata \\
HD 70958 & F3V & 6500 & -3.0 & 0.0 & F8 V Fe-1.3 CH-0.7 &    6294 &    4.16 &     1.0 &   -0.38 & 5.52\tablenotemark{x} & 0.2 & 3.3 \\
HD 39891 & F3V & 7000 & +0.0 & 0.0 & F4 V Fe-0.8 & 6714 & 3.76 & 1.9 & -0.30 &  5.61 & 0.10 & 1.5 \\
HD 137909 & F0p & 8000 & +0.3 & 0.1 & A8 V: SrCrEu       &    7624 &    3.99 &     2.0 &    0.50 & 7.00 & 0.20 & old \\
HD 8556 & F4V & 6750 & -1.0 & 0.0 & F5 V Fe-0.7 CH-0.3 & 6562 & 3.92 & 1.4 & -0.34 & 8.96 & 0.20 & 2.0 \\
HD 118889 & F1V & 7000 & +0.1 & 0.07 &  & (7000) & & & & 10.7 & 0.20 & 0.9 \\
HD 6767 & A3IV & 8000 & -1.0 & 0.0 & A6 V mA3 & 8077 & 4.11 & 1.7 & -0.15 & 12.1 & 0.20 & young \\
HD 127726 & A7Vn & 7750 & +0.1 & 0.05 & F0 Vn kA4mA3 comp & 7574 & 3.87 & 1.2 & -0.11 & 14.3 & 0.2 & young \\
HD 200499 & A5V & 8250 & -0.2 & 0.06 & A5 IV-V & 8084 & 3.97 & 1.8 & -0.02 & 14.5 & 0.30 & 0.5\tablenotemark{y} \\
HD 72462 & F0Vn & 7500 & +0.2 & 0.0 & A9 Vn & 7390 & 3.71 & 2.1 & 0.01 &  21.3 & 0.30 & old \\
HD 31925 & F3V+F9V & 6500 & -3.0 & 0.0 & F6 V Fe-1 CH-0.5 & 6357 & 3.72 & 1.3 & -0.46 &  21.6 & 0.50 & 2.1 \\
HD 100203 & F8V & 6000 & -1.0 & 0.0 & F6.5 V & 6151 & 3.95 & 1.6 & -0.29 & 22.3 & 0.81 & 5.1 \\
HD 46273\tablenotemark{e} & F2V & 6750 & -0.3 & 0.0 & F3 Vn & 6678 & 3.50 & 1.9 & -0.26 &  25.9 & 0.50 & 1.1 \\
HD 51733 & F3V & 7000 & +0.3 & 0.1 & F2 V & 6804 & 3.71 & 1.7 & -0.22 &  27.0 & 0.70 & 1.2 \\
HD 27710 & F2V & 7000 & -0.3 & 0.0 & F2 V & 6854 & 4.00 & 1.8 & -0.20 & 27.2 & 0.50 & 1.2 \\
HD 213235 & F2V & 7000 & +0.5 & 0.11 & kA7hF2mF5 III\tablenotemark{g} & 6856 & 3.48 & 4.0 & -0.21 &  37.0 & 0.70 & old \\
HD 13594 & F4V & 6750 & -0.5 & 0.0 & F5 V Fe-0.7 & 6626 & 4.10 & 1.1 & -0.21 &  37.4 & 0.90 & 2.4 \\
HD 661 & F2V+F6V & 7000 & -2.5 & 0.0 & F5 IV & 6861 & 3.69 & 2.7 & -0.11 & 46.5 & 0.70 & 0.9  \\ \hline
HD 120987 & F4V & 6500 & -0.5 & 0.1 & F5 V Fe-0.7 & 6473 & 3.77 & 1.4 & -0.27 & 50.1 & 1.00 & 1.6 \\
HD 110379 & F0V+FOV & 7500 & -0.5 & 0.21 & F2 V\tablenotemark{h}  & 6867\tablenotemark{i} &  3.65\tablenotemark{i} & 1.5\tablenotemark{i} & -0.20\tablenotemark{i} &   51.1 & 3.70 & 1.4 \\
HD 147365 & F3IV-V & 6750 & -3.0 & 0.0 & F4 V               &    6672 &    4.15 &     1.8 &   -0.09 & 53.6 & 2.00 & 0.6 \\
HD 80441 & F2V+F4V & 6750 & -0.3 & 0.0 & F5 V Fe-0.5 & 6558 & 4.11 & 1.8 & -0.32 & 66.9 & 1.40 & 2.2 \\
HD 173608 & F0Vn & 8000 & -1.5 & 0.0 &  & (8000) & & & &  75.5 & 2.50 & \nodata \\
HD 194943 & F3V & 7500 & -0.5 & 0.5 & F2 V               &    6771 &    3.53 &     2.0 &   -0.30 & 75.8 & 2.50 & 1.3 \\
HD 10453 & F5V+... & 6750 & -3.0 & 0.0 & F6 V Fe-1 CH-0.5   &    6457 &    4.04 &     1.4 &   -0.34 & 78.3 & 2.10 & 3.7 \\
HD 76644 & A7 & 8000 & -0.5 & 0.1 & A7 V(n)            &    7769 &    3.91 &     2.0 &    0.00 & 120 & 8.20 & 0.05\tablenotemark{y} \\
HD 8224 & F7V & 6250 & -0.5 & 0.0 & F8 V Fe-0.4 & 6269 & 4.18 & 1.0 & -0.11 & 128 & 2.60 & 6.2 \\
HD 11944 & F2V & 6750 & -1.0 & 0.0 & F4 V Fe-0.8 & 6675 & 4.06 & 1.5 & -0.34 &  132 & 2.40 & 2.0 \\
HD 99211 & A5V & 7750 & -0.2 & 0.0 & A7 V(n)            &    7805 &    3.86 &     2.0 &    0.04 & 136 & 5.30 & 0.7\tablenotemark{z}  \\
HD 129798 & F4V & 6750 & -0.5 & 0.0 &F4 V Fe-0.5 & 6716 & 3.88 & 1.4 & -0.11 & 153 & 3.60 & 1.7 \\
HD 191104 & F3V & 6500 & -0.5 & 0.1 & & (6500) & & & & 165 & 3.80 & 2.6 \\
HD 50635 & F0Vp & 7250 & +0.5 & 0.0 & F1 V(n) kA8mA8 & 7299 & 4.02 & 2.4 & -0.18 & 181 & 6.50 & young \\
HD 20631 & F3V & 7250 & -0.5 & 0.0 &  F2 V               &    6865 &    3.79 &     2.0 &   -0.24 & 264 & 7.2 & 1.4 \\
HD 17627 & F3/F5IV & 7250 & +0.5 & 0.1 & F3 V/F9 V & 6699\tablenotemark{h} & 3.92\tablenotemark{h} & 1.7\tablenotemark{h} & -0.22\tablenotemark{h} & 303 & 5.4 & 2.0 \\
HD 51199 & F2IV/V & 7250 & +0.2 & 0.0 & F1.5 V & 6780 & 3.62 & 1.6 & -0.09 & 339 & 11.6 & 1.5 \\
HD 91889 & F7V & 6250 & +0.0 & 0.1 & F8 V               &    6119 &    4.10 &     1.0 &   -0.14 & 354 & 14.4 & 7.4  \\
HD 119124 & F7.7V & 6250 & -0.2 & 0.0 & F8 V               &    6156 &    4.38 &     2.0 &   -0.20 & 444 & 17.6 & 5.5 \\ \hline
HD 16628 & A3V & 8500 & -0.3 & 0.0 & A3 V- & & & & & 2050 & 28.6 & \nodata \\
HD 142908 & F0IV & 7000 & -0.3 & 0.1 & F2 V & 6870 & 3.70 & 1.9 & -0.12 & 3910 & 94.3 & 1.3 \\
HD 61497 & A3IVn & 8750 & -0.3 & 0.2 & A7 Vn kA2mA2 & 7905\tablenotemark{i} & 3.65\tablenotemark{i} & 2.0\tablenotemark{i} & -0.84\tablenotemark{i} &  3980 & 54.7 & young \\
HD 77190 & A8Vn & 7750 & -3.0 & 0.0 & A8 V(n) & 7703 & 3.94 & 1.4 & 0.12 & 6100 & 103.9 & \nodata \\
HD 196885 & F8IV & 6500 & -3.0 & 0.1 & F8 IV-V & 6221 & 4.26 & 1.0 & 0.17 & 6330 & 191.9 & 8.4 \\
HD 111066 & F8V & 6250 & -0.5 & 0.06 & F8+ V & 6136 & 4.21 & 1.0 & -0.11 & 6970 & 160.3 & 6.1 \\ 
\enddata
\tablenotetext{a}{Teff and metallicity of the best-fit Kurucz models
(see Appendix A for discussion).}
\tablenotetext{b}{Extinction, a free parameter,
required to fit available
photometry for best-fit Kurucz model (see Appendix~A).}
\tablenotetext{c}{Spectral type
and physical properties
obtained 
following the technique of
Gray \& Corbally (see Appendix~A).
The label ``comp'' indicates that the best explanation for the
appearance of the spectrum is a composite of two stars. Where 
there is no spectral type from G\&C we list an effective
temperature, in parentheses, interpolated from the Simbad
spectral type (see Appendix A).
No Str\"{o}mgren photometry is available for
HD~16628, so there is no G\&C temperature for
that system.}
\tablenotetext{d}{Spectral classifications are in agreement
with \citet{hummel}, who found that the primary is F9III (e.g., G\&C) and
the secondary is A5V (e.g., Simbad).}
\tablenotetext{e}{Quadruple system: HD~17094 \citep{richichi00}, HD~46273 \citep{nord04}}
\tablenotetext{f}{HD~118216: Strong reversals in K\&H lines: RS CVn star. HD~83808: spectrum clearly composite with weak Ca K line.}
\tablenotetext{g}{Possible low metallicity for the Am star.}
\tablenotetext{h}{HD~110379: Spectral type is composite of HD110379/80, but fit is only for component~A.
HD~17627: Only northern component (F3 V) fit.}
\tablenotetext{i}{Poor fit due to large photometry residuals.}
\tablecomments{Targets
listed in order of
increasing separation. Two horizontal
lines divide the sample into its three
separation bins: $<$3~AU; 3--50~AU; and
$>$50~AU.
Additionally, the six systems with extremely wide ($>$500~AU)
separations
are separated by a horizontal line.
All separations are from CCDM \citep{ccdm}, SB9 \citep{sb9}, and/or WDS \citep{wds}, except
as indicated here. Some of the non-CCDM/SB9 separations
require additional basic information such as
parallax or assumption of stellar masses,
generally from Simbad for the parallaxes and
\citet{lang}
for stellar masses.
References for non-CCDM/SB9 separations:
j: \citet{budding04};
k: \citet{margoni92};
l: \citet{prieur03};
m: \citet{sb9};
n: \citet{strass93};
o: \citet{mayor87};
p: \citet{paunzen98};
q: \citet{giuricin84};
r: \citet{vennes98};
s: \citet{richichi02};
t: \citet{abt76};
u: \citet{pourbaix00};
v: \citet{soder99};
w: \citet{roberts05};
x: \citet{mcalister93}.
Ages are from \citet{nord04} unless otherwise
indicated, as follows:
y: I.\ Song, pers.\ comm.;
z: \citet{song01}.
Ages for 14~systems are not available,
as indicated; ``young'' and ``old'' are defined in the text.
}
\end{deluxetable}

\begin{deluxetable}{llll}
\tabletypesize{\scriptsize} 
\tablecaption{Observing log\label{obsinfo}} 
\tablewidth{0pt}
\tablehead{ \colhead{Name}
& \colhead{Int.\ time}
& \colhead{Int.\ time} 
& \colhead{AORKey\tablenotemark{a}} \\
\colhead{}     
& \colhead{24~$\mu$m} 
& \colhead{70~$\mu$m} 
& \colhead{} \\         
\colhead{}  
& \colhead{(sec)}  
& \colhead{(sec)}  
& \colhead{}}  
\startdata
HD 207098 & 48 & 252\tablenotemark{b} & 4227584  \\
HD 106112 & 48 & 231 & 4228864  \\
HD 118216 & 48 & \nodata & 4232704  \\
HD 150682 & 48 & 1091 & 4232192  \\
HD 29140 & 48 & 1091 & 4232448   \\
HD 199532 & 48 & 1091 & 4232960  \\
HD 119756 & 48 & 231 & 4228352 \\
HD 16920 & 48 & 1091 & 4233216  \\
HD 20320 & 48 & 440 & 4230400   \\
HD 204188 & 48 & 1091 & 4231936   \\
HD 83808 & 48 & 881 & 4231680   \\
HD 88215 & 48 & 440 & 4229632  \\
HD 178449 & 48 & 881 & 4231424   \\
HD 13161 & 48 & 545 & 4230912   \\
HD 217792 & 48 & 231 & 4229120  \\
HD 11636 & 48 & 252\tablenotemark{b} & 4228096  \\
HD 151613 & 48 & 231, 440 & 4228608, 12635904   \\
HD 99028 & 48 & 440 & 4229888  \\
HD 17094 & 48 & 440 & 4230656  \\
HD 32537 & 48 & 231, 440 & 4229376, 12635136   \\
HD 17206 & 48 & 84\tablenotemark{b} & 8932096 \\ \hline
HD 80671 & 48 & 650 & 4233984   \\
HD 56986 & 48 & 101\tablenotemark{b} & 4233472   \\
HD 26690 & 48 & 1091 & 4235264   \\
HD 10009 & 48 & \nodata & 4243712  \\
HD 195725 & 48 & 650 & 4231168  \\
HD 95698 & 48 & 1091 & 4236544   \\
HD 70958 & 48 & 440, 881 & 4230144, 12636416  \\
HD 39891 & 48 & 1091 & 4237312   \\
HD 137909 & 48 & 545 & 4233728   \\
HD 8556 & 48 & 1091 & 4237056   \\
HD 118889 & 48 & \nodata & 4237824  \\
HD 6767 & 48 & 1091 & 4234752  \\
HD 127726 & 48 & 1091 & 4244480, 12636160   \\
HD 200499 & 48 & 1091 & 4235776  \\
HD 72462 & 48 & \nodata & 4239104  \\
HD 31925 & 48 & 1091 & 4235520  \\
HD 100203 & 48 & 1091 & 4234240   \\
HD 46273 & 48 & 1091 & 4236288  \\
HD 51733 & 48 & 1091 & 4235008  \\
HD 27710 & 48 & 1091 & 4236800  \\
HD 213235 & 48 & \nodata & 4238080  \\
HD 13594 & 48 & 1091 & 4236032  \\
HD 661 & 48 & \nodata & 4238592    \\ \hline
HD 120987 & 48 & \nodata & 4244736  \\
HD 110379 & 48 & 132 & 4239616  \\
HD 147365 & 48 & 336, 650 & 4240128, 12635392  \\
HD 80441 & 48 & \nodata & 4243968   \\
HD 173608 & 48 & 336 & 4240384   \\
HD 194943 & 48 & 336 & 4241920   \\
HD 10453 & 48 & \nodata & 4242688  \\
HD 76644 & 48 & 132 & 4239360   \\
HD 8224 & 48 & \nodata & 4244224   \\
HD 11944 & 48 & \nodata & 4243456  \\
HD 99211 & 48 & 132 & 4239872   \\
HD 129798 & 48 & 336 & 4242176   \\
HD 191104 & 48 & \nodata & 4242944   \\
HD 50635 & 48 & 440, 650 & 4241152, 12635648  \\
HD 20631 & 48 & 336 & 4241664   \\
HD 17627 & 48 & \nodata & 4244992   \\
HD 51199 & 48 & 440 & 4240896  \\
HD 91889 & 48 & 650 & 4241408    \\
HD 119124 & 48 & 440 & 4240640    \\ \hline
HD 16628 & 48 & 1091 & 4237568, 12634880  \\
HD 142908 & 48 & 881 & 4234496   \\
HD 61497 & 48 & \nodata & 8934144    \\
HD 77190 & 48 & \nodata & 4238336  \\
HD 196885 & 48 & 440 & 4242432  \\
HD 111066 & 48 & \nodata  &4245248   \\
\enddata
\tablenotetext{a}{Further details of each observation,
including pointing and time and date of observation, can
be queried from the {\it Spitzer} Data Archive at
the {\it Spitzer} Science Center.}
\tablenotetext{b}{Narrow field observation. The calibration factor
is 17\% higher for these observations than for default
scale 70~\micron\ observations.}
\tablecomments{As in Table~\ref{targetinfo}, targets are
listed in order of increasing separation, with horizontal
lines indicating subsamples.}
\end{deluxetable} 

\begin{deluxetable}{lccccccccccc}
\tabletypesize{\scriptsize}
\tablecaption{Photospheric predictions and photometry for all sources\label{photom}}
\tablewidth{0pt}
\tablehead{
\colhead{Name} 
& \colhead{V\tablenotemark{a}} 
& \colhead{K\tablenotemark{a}}
& \colhead{F24}
& \colhead{P24}
& \colhead{R24}
& \colhead{$\chi_{24}$}
& \colhead{F70}
& \colhead{P70} 
& \colhead{R70}
& \colhead{$\chi_{70}$}
& \colhead{F870\tablenotemark{b}} \\
\colhead{}
& \colhead{(mag)}
& \colhead{(mag)}
& \colhead{(mJy)}
& \colhead{(mJy)}
& \colhead{} & \colhead{}
& \colhead{(mJy)}
& \colhead{(mJy)}
& \colhead{} & \colhead{}
& \colhead{(mJy)}}
\startdata
HD 207098 & 2.85 & 2.06$\pm$0.13 & 1026$\pm$41 & 1004 & 1.02 & 0.52 & 119$\pm$12 & 112 & 1.07 & 0.63 & \\   
HD 106112 & 5.14 & 4.38$\pm$0.12 & 124$\pm$5 & 124 & 1.00 & -0.03 & 14$\pm$4 & 13 & 1.02 & 0.07 & $<$19 \\  
HD 118216 & 4.91 & 3.62$\pm$0.16 & 361$\pm$14 & 245 & 1.47 & 8.03 & \nodata & 27 & \nodata & \nodata & $<$35 \\  
HD 150682 & 5.92 & 4.89$\pm$0.01 & 78$\pm$3 & 77 & 1.01 & 0.36 & 30$\pm$3 & 8 & 3.58 & 8.61 & $<$92 \\  
HD 29140 & 5.13 & 3.65$\pm$0.13 & 217$\pm$9 & 212 & 1.02 & 0.60 & 28$\pm$3 & 23 & 1.20 & 1.39 & $<$25 \\
HD 199532 & 4.25 & 3.84$\pm$0.11 & 268$\pm$11 & 258 & 1.04 & 0.91 & 41$\pm$3 & 29 & 1.41 & 3.59 & \\   
HD 119756 & 5.21 & 4.24$\pm$0.10 & 340$\pm$14 & 341 & 1.00 & -0.03 & 16$\pm$5 & 38 & 0.42 & -4.83 & \\   
HD 16920 & 4.23 & 3.35$\pm$0.14 & 153$\pm$6 & 128 & 1.20 & 4.18 & 17$\pm$3 & 14 & 1.22 & 1.22 & \\   
HD 20320 & 4.80 & 4.23$\pm$0.11 & 158$\pm$6 & 143 & 1.11 & 2.47 & 103$\pm$8 & 15 & 6.66 & 10.56 & \\   
HD 204188 & 6.08 & 5.49$\pm$0.03 & 42$\pm$2 & 46 & 0.91 & -2.54 & \nodata$\pm$6 & 5 & \nodata & \nodata & \\   
HD 83808 & 3.52 & 2.58$\pm$0.15 & 822$\pm$33 & 670 & 1.23 & 4.64 & 94$\pm$8 & 72 & 1.30 & 2.91 & \\
HD 88215 & 5.30 & 4.40$\pm$0.11 & 139$\pm$6 & 125 & 1.11 & 2.46 & 22$\pm$2 & 14 & 1.61 & 4.38 & \\   
HD 178449 & 5.20 & 4.19$\pm$0.12 & 153$\pm$6 & 157 & 0.97 & -0.70 & 14$\pm$1 & 17 & 0.82 & -2.45 & \\   
HD 13161 & 3.00 & 2.69$\pm$0.18 & 756$\pm$30 & 587 & 1.29 & 5.65 & 643$\pm$51 & 65 & 9.88 & 11.23 & \\   
HD 217792 & 5.12 & 4.36$\pm$0.10 & 137$\pm$5 & 121 & 1.13 & 2.83 & 32$\pm$3 & 14 & 2.37 & 6.71 & \\   
HD 11636 & 2.64 & 2.29$\pm$0.12 & 817$\pm$33 & 830 & 0.98 & -0.41 & 85$\pm$12 & 91 & 0.93 & -0.52 & $<$26 \\  
HD 151613 & 4.84 & 3.81$\pm$0.10 & 212$\pm$9 & 220 & 0.97 & -0.90 & 35$\pm$4 & 24 & 1.45 & 2.97 & $<$19 \\  
HD 99028 & 4.00 & 2.82$\pm$0.13 & 505$\pm$20 & 501 & 1.01 & 0.23 & 50$\pm$4 & 54 & 0.93 & -0.93 & \\   
HD 17094 & 4.27 & 3.45$\pm$0.13 & 299$\pm$12 & 297 & 1.01 & 0.14 & 71$\pm$6 & 32 & 2.20 & 6.75 & \\   
HD 32537 & 4.98 & 4.11$\pm$0.10 & 164$\pm$7 & 171 & 0.96 & -1.04 & 17$\pm$4 & 19 & 0.91 & -0.41 & $<$33 \\  
HD 17206 & 4.47 & 3.22$\pm$0.17 & 336$\pm$13 & 343 & 0.98 & -0.53 & 43$\pm$16 & 37 & 1.17 & 0.38 & \\   \hline
HD 80671 & 5.38 & 4.42$\pm$0.10 & 125$\pm$5 & 125 & 1.00 & 0.02 & 18$\pm$2 & 14 & 1.30 & 2.69 & \\   
HD 56986 & 3.50 & 2.62$\pm$0.31 & 684$\pm$27 & 632 & 1.08 & 1.89 & 67$\pm$10 & 69 & 0.98 & -0.15 & \\   
HD 26690 & 5.29 & 4.47$\pm$0.15 & 129$\pm$5 & 117 & 1.10 & 2.37 & 12$\pm$1 & 13 & 0.96 & -0.40 & \\   
HD 10009 & 6.24 & 4.91$\pm$0.01 & 75$\pm$3 & 76 & 0.99 & -0.36 & \nodata & 8 & \nodata & \nodata & \\   
HD 195725 & 4.21 & 3.69$\pm$0.16 & 221$\pm$9 & 255 & 0.86 & -3.93 & 24$\pm$2 & 28 & 0.88 & -1.67 & $<$22 \\  
HD 95698 & 6.21 & 5.42$\pm$0.02 & 55$\pm$2 & 49 & 1.12 & 2.63 & 137$\pm$11 & 5 & 25.56 & 11.84 & \\   
HD 70958 & 5.61 & 4.49$\pm$0.07 & 127$\pm$5 & 125 & 1.01 & 0.33 & 18$\pm$3 & 14 & 1.27 & 1.41 & $<$25 \\  
HD 39891 & 6.34 & 5.38$\pm$0.01 & 51$\pm$2 & 49 & 1.06 & 1.33 & \nodata$\pm$5 & 5 & \nodata & \nodata & \\   
HD 137909 & 3.66 & 3.48$\pm$0.16 & 341$\pm$14 & 416 & 0.82 & -5.50 & 40$\pm$3 & 45 & 0.89 & -1.53 & $<$50 \\  
HD 8556 & 5.92 & 4.91$\pm$0.01 & 78$\pm$3 & 78 & 0.99 & -0.14 & 12$\pm$3 & 9 & 1.35 & 1.01 & \\   
HD 118889 & 5.57 & 4.77$\pm$0.02 & 92$\pm$4 & 90 & 1.03 & 0.65 & \nodata & 10 & \nodata & \nodata & \\   
HD 6767 & 5.21 & 4.80$\pm$0.01 & 86$\pm$3 & 86 & 1.00 & -0.08 & 8$\pm$2 & 9 & 0.87 & -0.69 & \\   
HD 127726 & 6.00 & 5.39$\pm$0.01 & 49$\pm$2 & 49 & 1.00 & 0.07 & 16$\pm$2 & 5 & 3.10 & 4.94 & \\
HD 200499 & 4.82 & 4.43$\pm$0.13 & 122$\pm$5 & 111 & 1.10 & 2.18 & 17$\pm$3 & 12 & 1.45 & 1.98 & \\   
HD 72462 & 6.38 & 5.68$\pm$0.02 & 38$\pm$2 & 37 & 1.03 & 0.69 & \nodata & 4 & \nodata & \nodata & \\   
HD 31925 & 5.65 & 4.50$\pm$0.03 & 115$\pm$5 & 123 & 0.93 & -1.81 & 17$\pm$1 & 14 & 1.25 & 2.36 & \\   
HD 100203 & 5.46 & 4.16$\pm$0.11 & 154$\pm$6 & 180 & 0.85 & -4.32 & 23$\pm$2 & 20 & 1.17 & 1.80 & \\
HD 46273 & 5.28 & 4.34$\pm$0.11 & 134$\pm$5 & 138 & 0.97 & -0.87 & 29$\pm$5 & 15 & 1.94 & 2.71 & \\   
HD 51733 & 5.45 & 4.50$\pm$0.11 & 115$\pm$5 & 105 & 1.10 & 2.23 & 18$\pm$4 & 11 & 1.54 & 1.76 & \\   
HD 27710 & 6.08 & 5.19$\pm$0.01 & 59$\pm$2 & 59 & 1.01 & 0.14 & 11$\pm$4 & 6 & 1.78 & 1.32 & \\   
HD 213235 & 5.51 & 4.59$\pm$0.08 & 101$\pm$4 & 106 & 0.96 & -1.18 & \nodata & 11 & \nodata & \nodata & \\   
HD 13594 & 6.05 & 4.98$\pm$0.02 & 73$\pm$3 & 72 & 1.02 & 0.46 & 5$\pm$1 & 8 & 0.66 & -3.04 & \\   
HD 661 & 6.66 & 5.70$\pm$0.01 & 37$\pm$1 & 37 & 0.98 & -0.64 & \nodata & 4 & \nodata & \nodata & \\    \hline
HD 120987 & 5.53 & 4.41$\pm$0.09 & 122$\pm$5 & 123 & 0.99 & -0.27 & \nodata & 13 & \nodata & \nodata & \\   
HD 110379 & 2.74 & 1.88$\pm$0.14 & 1367$\pm$55 & 1287 & 1.06 & 1.46 & 134$\pm$19 & 140 & 0.96 & -0.32 & \\   
HD 147365 & 5.48 & 4.39$\pm$0.15 & 123$\pm$5 & 131 & 0.94 & -1.61 & 13$\pm$2 & 15 & 0.87 & -0.79 & \\   
HD 80441 & 6.12 & 5.05$\pm$0.01 & 66$\pm$3 & 69 & 0.96 & -1.10 & \nodata & 7 & \nodata & \nodata & \\   
HD 173608 & 4.59 & 4.16$\pm$0.08 & 147$\pm$6 & 162 & 0.90 & -2.65 & 25$\pm$2 & 18 & 1.40 & 3.35 & \\   
HD 194943 & 4.77 & 3.70$\pm$0.11 & 216$\pm$9 & 244 & 0.89 & -3.18 & 32$\pm$5 & 27 & 1.19 & 1.02 & \\   
HD 10453 & 5.75 & 4.69$\pm$0.01 & 101$\pm$4 & 98 & 1.03 & 0.75 & \nodata & 11 & \nodata & \nodata & \\   
HD 76644 & 3.12 & 2.70$\pm$0.13 & 635$\pm$25 & 602 & 1.05 & 1.29 & 79$\pm$7 & 66 & 1.20 & 1.81 & \\   
HD 8224 & 7.00 & 5.70$\pm$0.02 & 38$\pm$2 & 38 & 1.00 & 0.01 & \nodata & 4 & \nodata & \nodata & \\   
HD 11944 & 6.43 & 5.45$\pm$0.02 & 46$\pm$2 & 49 & 0.94 & -1.60 & \nodata & 5 & \nodata & \nodata & \\   
HD 99211 & 4.06 & 3.53$\pm$0.29 & 276$\pm$11 & 262 & 1.05 & 1.27 & 55$\pm$8 & 28 & 1.92 & 3.47 & \\   
HD 129798 & 6.24 & 4.68$\pm$0.05 & 60$\pm$2 & 99 & 0.61 & -15.70 & \nodata$\pm$5 & 11 & \nodata & \nodata & $<$30 \\  
HD 191104 & 6.43 & 5.22$\pm$0.01 & 64$\pm$3 & 57 & 1.13 & 2.74 & \nodata & 6 & \nodata & \nodata & $<$44 \\  
HD 50635 & 4.73 & 3.73$\pm$0.20 & 219$\pm$9 & 242 & 0.90 & -2.62 & 23$\pm$4 & 27 & 0.88 & -0.85 & $<$34 \\  
HD 20631 & 5.72 & 4.76$\pm$0.03 & 92$\pm$4 & 92 & 1.00 & -0.10 & 22$\pm$4 & 10 & 2.19 & 2.94 & \\   
HD 17627 & 6.66 & 5.52$\pm$0.04 & 40$\pm$2 & 42 & 0.94 & -1.43 & \nodata & 5 & \nodata & \nodata & \\   
HD 51199 & 4.66 & 3.84$\pm$0.18 & 246$\pm$10\tablenotemark{f} & 212 & 1.16 & 3.47 & 39$\pm$5 & 23 & 1.69 & 3.27 & \\   
HD 91889 & 5.71 & 4.35$\pm$0.07 & 126$\pm$5 & 140 & 0.90 & -2.79 & 6$\pm$4 & 16 & 0.36 & -2.49 & \\   
HD 119124 & 6.31 & 4.87$\pm$0.02 & 84$\pm$3\tablenotemark{f} & 85 & 0.99 & -0.23 & 74$\pm$7 & 9 & 7.81 & 9.40 & \\   \hline
HD 16628 & 5.30 & 5.01$\pm$0.02 & 85$\pm$3\tablenotemark{f} & 69 & 1.23 & 4.58 & 42$\pm$4 & 7 & 5.67 & 8.00 & \\   
HD 142908\tablenotemark{e} & 5.43 & 5.50$\pm$0.09 & 112$\pm$4 & 117 & 0.96 & -1.12 & 33$\pm$4 & 13 & 2.59 & 5.80 & \\   
HD 61497\tablenotemark{e} & 4.93 & 4.64$\pm$0.08 & 98$\pm$4 & 103 & 0.95 & -1.26 & \nodata & 11 & \nodata & \nodata & \\   
HD 77190\tablenotemark{e} & 6.07 & 5.43$\pm$0.01 & 47$\pm$2\tablenotemark{f} & 49 & 0.97 & -0.76 & \nodata & 5 & \nodata & \nodata & \\   
HD 196885\tablenotemark{e} & 6.39 & 5.10$\pm$0.01 & 65$\pm$3 & 67 & 0.97 & -0.79 & 6$\pm$3 & 8 & 0.84 & -0.35 & \\   
HD 111066\tablenotemark{e} & 6.83 & 5.53$\pm$0.01 & 42$\pm$2\tablenotemark{f} & 45 & 0.95 & -1.45 & \nodata & 5 & \nodata & \nodata & \\ 
\enddata
\tablenotetext{a}{V~magnitudes from Hipparcos, with typical
errors 0.01~mag \citep{perryman};
K~magnitudes are ``super-K,''
which is 
a higher SNR version of the 2MASS K~magnitude created by combining
the 2MASS J,~H~and K~magnitudes \citep{2mass}, suitably corrected for the expected colors
of our target stars \citep{tokunaga}.}
\tablenotetext{b}{Three sigma upper limits,
where available. None of the 13~sources observed
at 870~\micron\ were detected.}
\tablenotetext{c}{No reliable error available from the 2MASS catalog.}
\tablenotetext{d}{Not in the 2MASS catalog.}
\tablenotetext{e}{These systems are resolved at all wavelengths.
The data presented here for these systems is only for
the primary.}
\tablenotetext{f}{The F24 fluxes and errors presented here are system
integrated, but the secondary is resolvable at 24~\micron\
and within the field of view.
The fluxes for the secondaries only are as follows, in mJy
(where the errors include the standard calibration
error of 4\%):
HD~51199, 33$\pm$1;
HD~119124, 13$\pm$1;
HD~16628, 5$\pm$1;
HD~77190, 19$\pm$1 and
HD~111066, 8$\pm$1.}
\tablecomments{Measured (``F'') and predicted
(``P'') fluxes for all systems are listed as well
as significances ($\chi$).
All numbers represented system-integrated fluxes
and magnitudes except for the five indicated systems.
All measurements are subject to both photometric (measurement)
error and a uniform calibration
uncertainty of 4\% at 24~\micron\ and 8\% at 70~\micron. These
two sources of error are 
RSS-combined to calculate the total errors presented here.
At 70~\micron, a number of 
systems were not targeted; these are indicated
by ``\ldots'' in the F$_{70}$~column. Three 
systems were targeted at 70~\micron\ but not
detected; these are indicated by
``\ldots$\pm\sigma$'' where $\sigma$ is the 
total error as defined above and including
the noise in the image as the photometric error
contribution.
As in Table~\ref{targetinfo}, targets are
listed in order of increasing separation, with horizontal
lines indicating subsamples.}
\end{deluxetable}

\begin{deluxetable}{lcccc|cccc|cccc}
\tabletypesize{\scriptsize}
\rotate
\tablecaption{Excesses detected at 24 and 70 \micron\ and properties of detected dust\label{excesstable}}
\tablewidth{0pt}
\tablehead{
\colhead{Name} &
\colhead{R24} &
\colhead{$\chi_{24}$} & 
\colhead{R70} & 
\colhead{$\chi_{70}$} & 
\colhead{Binary sep.}
& \colhead{e} & \colhead{Stellar masses}
& \colhead{Refs.\tablenotemark{a}}
& \colhead{Dust temp.\tablenotemark{b}} & \colhead{Dust dist.\tablenotemark{b}} &
\colhead{Dynamical} &
\colhead{Frac.\ Lum.\tablenotemark{b}} \\
\colhead{} & 
\colhead{} & 
\colhead{} & 
\colhead{} & 
\colhead{} & 
\colhead{(AU)} & 
\colhead{} & 
\colhead{($M_\odot$)} & 
\colhead{} &  
\colhead{(K)} & 
\colhead{(AU)} & 
\colhead{state\tablenotemark{c}} & 
\colhead{$\times10^{-5}$}} 
\startdata
HD 118216 & 1.47 & 8.03 & \nodata & \nodata & 0.04 & 0.0 & 1.5, 0.8 & 1          & $>$50 & $<$163 & sb & 200 \\
HD 13161 & 1.29\tablenotemark{f} & 5.65\tablenotemark{f} & 9.88\tablenotemark{f} & 11.23\tablenotemark{f} & 0.3 & 0.43 & 3.5, 1.4 & 2           & 93 & 81 & sb & 3.2 \\
HD 83808 & 1.23\tablenotemark{f} & 4.64\tablenotemark{f} & 1.30\tablenotemark{f} & 2.91\tablenotemark{f} & 0.21 & 0.0 & \nodata & 3            & 815 & 0.85 & sb & 46 \\ 
HD 16628\tablenotemark{d} & 1.23\tablenotemark{f} & 4.58\tablenotemark{f} & 5.67\tablenotemark{f} & 8.00\tablenotemark{f} & 2050 & \nodata & \nodata & \nodata & 103 &  42 & ss & 1.9 \\
HD 16920 & 1.20 & 4.18  & 1.22 & 1.22  & 0.10 &  \nodata  & \nodata  & \nodata     & $>$260 & $<$4.8 & sb & 2.9 \\
HD 51199 & 1.16\tablenotemark{f} & 3.47\tablenotemark{f} & 1.69\tablenotemark{f} & 3.27\tablenotemark{f} & 339 & \nodata & \nodata & \nodata & 188 & 6.7 & ss & 1.8 \\ \hline
HD 95698\tablenotemark{d} & 1.12\tablenotemark{g} & 2.63\tablenotemark{g} & 25.56 & 11.84 & 5.31 & \nodata & \nodata & \nodata  & 65 & 51 & sb & 9.3 \\
HD 13161 & 1.29\tablenotemark{f} & 5.65\tablenotemark{f} & 9.88\tablenotemark{f} & 11.23\tablenotemark{f} & 0.3 & 0.43 & 3.5, 1.4 & 2           & 93 & 81 & sb & 3.2 \\
HD 119124 & 0.99 & -0.23 & 7.81 & 9.40 & 444 & \nodata & \nodata & \nodata & 81 & 16 & ss & 5.8 \\
HD 20320 & 1.11\tablenotemark{g} & 2.47\tablenotemark{g} & 6.66 & 10.56 & 0.18 & 0.14 & \nodata & 3           & 86 & 38 & sb & 2.3 \\
HD 16628\tablenotemark{d} & 1.23\tablenotemark{f} & 4.58\tablenotemark{f} & 5.67\tablenotemark{f} & 8.00\tablenotemark{f} & 2050 & \nodata & \nodata & \nodata & 103 &  42 & ss & 1.9 \\
HD 150682 & 1.01 & 0.36 & 3.58 & 8.61 & 0.04  & 0.0 & \nodata & 4          & 102 & 19 & sb & 2.2 \\
HD 127726\tablenotemark{e} & 1.00 & 0.07 & 3.10 & 4.94 & 14.3 & 0.16 & 6.6, 6.9 & 3,5      & 108 & 28 & u & 1.3 \\
HD 142908 & 0.96 & -1.12 & 2.59 & 5.80 & 3910 & \nodata & \nodata & \nodata & 115 & 19 & ss & 1.5 \\
HD 217792 & 1.13\tablenotemark{g} & 2.83\tablenotemark{g} & 2.37 & 6.71 & 0.57 & 0.53 & \nodata & 3          & 127 & 12 & sb & 1.3 \\
HD 17094 & 1.01 & 0.14 & 2.20 & 6.75 & 2.58 & \nodata & \nodata & \nodata  & 129 & 17 & sb & 1.1 \\
HD 20631 & 1.00 & -0.10 & 2.19 & 2.94 & 264 & \nodata & \nodata & \nodata & 128 & 12 & ss & 1.4 \\
HD 46273 & 0.97 & -0.87 & 1.94 & 2.71 & 25.9  & 0.23 & \nodata & 6        & 139 & 16 & u & 1.4 \\
HD 99211 & 1.05 & 1.27 & 1.92 & 3.47 & 136 & \nodata & \nodata & \nodata & 146 & 13 & ss & 1.0 \\
HD 51199 & 1.16\tablenotemark{f} & 3.47\tablenotemark{f} & 1.69\tablenotemark{f} & 3.27\tablenotemark{f} & 339 & \nodata & \nodata & \nodata & 188 & 6.7 & ss & 1.8 \\
HD 88215 & 1.11\tablenotemark{g} & 2.46\tablenotemark{g} & 1.61 & 4.38 & 0.19 & \nodata & \nodata & \nodata   & 180 & 5.2 & sb & 1.5 \\
HD 151613 & 0.97 & -0.90 & 1.45 & 2.97 & 1.07 & \nodata & \nodata & \nodata & 197 & 5.3 & sb & 1.5 \\
HD 199532 & 1.04 & 0.91 & 1.41 & 3.59 & 0.05 & \nodata & \nodata & \nodata & 219 & 9.8 & sb & 1.2 \\
HD 173608\tablenotemark{d} & 0.90 & -2.65 & 1.40 & 3.35 & 75.5 & \nodata & \nodata & \nodata & 203 & 6.4 & ss & 0.9 \\
HD 80671 & 1.00 & 0.02 & 1.30481 & 2.69 & 3.35 & 0.50 & \nodata  & 6          & 266 & 2.9 & u & 2.0 \\
HD 83808 & 1.23\tablenotemark{f} & 4.64\tablenotemark{f} & 1.30\tablenotemark{f} & 2.91\tablenotemark{f} & 0.21 & 0.0 & \nodata & 3            & 815 & 0.85 & sb & 46 \\
\enddata
\tablenotetext{a}{References for binary eccentricities
and stellar masses, where available: (1) \citet{strass93}; (2) \citet{pourbaix00};
(3) \citet{abt05};
(4) \citet{mayor87}; (5) \citet{heintz};
(6) \citet{soder99}}
\tablenotetext{b}{Maximum temperatures, minimum
dust distances, and maximum fractional
luminosities. For HD~118216 and HD~16920,
minimum temperatures and maximum distances are given
(see text
for discussion).}
\tablenotetext{c}{Dynamical state of the derived
dust location:
``ss'' means the dust is in a stable location around
a single star (that is, circumstellar),
and ``sb'' means a stable circumbinary location.
The code ``u'' means unstable (see Section~\ref{unstable}).}
\tablenotetext{d}{System has no age estimate and therefore
could be young.}
\tablenotetext{e}{System is young.}
\tablenotetext{f}{System formally has excess emission at both
24~and 70~\micron.}
\tablenotetext{g}{Not a formal 24~\micron\ excess because
R24 is close to but does not exceed the threshold value,
although $\chi_{24}$ is $>$2.0.
Nevertheless, this is likely a 24~\micron\
excess, detected at a level that is not statistically significant.}
\tablecomments{The
first group (first six lines) are systems with 24~\micron\ excess
(R24~$\geq$~1.15 and $\chi_{24}\geq2.0$), in order of decreasing R24;
the second group (next 20~lines) are systems with 70~\micron\ excess 
(R70~$\geq$~1.30 and $\chi_{70}\geq2.0$),
in order of decreasing R70.
There are four systems listed twice, once in the
top part of the table and once in the bottom part,
since these systems have excesses at both
wavelengths.
The left columns are based on our measurements;
the middle columns contain literature values for 
these binary systems (separations from Table~\ref{targetinfo}), where
available; 
and the right columns are 
results of our modeling.}
\end{deluxetable}


\clearpage



\thispagestyle{empty}

\thispagestyle{empty}
\clearpage

\end{document}